\documentclass[twocolumn]{aastex62}

\received{November 30, 2019}
\revised{February 18, 2020}
\accepted{February 25, 2020}
\submitjournal{ApJ}
\shorttitle{Pulse Features of BATSE Short GRBs }
\shortauthors{X.-J., Li et al.}
\usepackage{amsmath}
\usepackage{graphicx}
\usepackage{float}
\usepackage{multirow}
\usepackage{subfigure}
\begin{document}

\title{Properties of Short GRB Pulses in the Fourth BATSE Catalog: Implications for Structure and Evolution of the Jetted Outflows
}
\author{Xiu-Juan Li}
\author{Zhi-Bin Zhang*}
\author{Chuan-Tao Zhang}
\author{Kai Zhang}
\author{Ying Zhang}
\author{Xiao-Fei Dong}
\affiliation{College of Physics and Engineering, Qufu Normal University, Qufu 273165, People's Republic of China;*Email:astrophy0817@163.com }
\begin{abstract}
Considering the shortage of comparative studies on pulse temporal characteristics between single-peaked short gamma-ray bursts (sGRBs) and double-peaked sGRBs,
we examine the pulse properties of a sample of 100 BATSE sGRBs using the BATSE Time-Tagged Event (TTE) data with a 5-ms resolution. 243 individual pulses of the
single-peaked, double-peaked and triple-peaked sGRBs are fitted to get their statistical properties such as the pulse asymmetry, amplitude, peak time, and pulse
width, etc. For the double-peaked sGRBs, according to the overlapping ratio between two adjacent peaks, we first define two kinds of double-peaked sGRBs as
M-loose and M-tight types and find that most of the first pulses are similar to the single-peaked ones. We study the dependence of the Full Width at Half
Maximum (FWHM) on the photon energy among different energy bands. Interestingly, it is found that a power-law relation with an index of -0.4 does exist between
the FWHM and the photon energy for the single- and double-peaked sGRBs. More interestingly, we notice that the power-law relation with a positive index also
exist for several special short bursts. In view of the three typical timescales of pulses, namely the angular spreading timescale, the dynamic timescale, and
the cooling timescale, we propose that the diverse power-law indexes together with the profile evolution of GRB pulse can be used as an effective probe to
diagnose the structure and evolution of the relativistically jetted outflows.
\end{abstract}

\section{Introduction} \label{sec:intro}
Gamma-ray bursts (GRBs) had been studied about forty years since the first burst reported for GRB 700822 \citep{1973ApJ...182...L85}, the physical processes and
radiation mechanisms of prompt emissions still have a lot of large debates. Study of the prompt emission light-curve and the spectral evolutions can provide a
good probe to learn about the nature of these problems. Many authors have investigated the pulse properties via different methods in the past decades
years \citep[e.g.,][]{1994ApJS...92...637,1999AAS...138...419,2002AA...385...377,1996ApJ...459...393,2005ApJ...627...324,Norrisetal2011,2011ApJ...740...104,2014ApJ...783...88,
2015ApJ...815...134,2018ApJ...855...101,2018ApJ...863...77}. GRB temporal profiles are usually very various, irregular and complicated. There are no two bursts
with the same temporal and spectral evolution found ever. Pulses as the basic radiative units of prompt gamma-ray emissions, contain the key information about the
physical mechanisms and the environments in which GRBs are generated \citep{1996ApJ...459...393}. Since the GRB light curves exhibit very complex structures due
to the overlapping of different neighbor pulses, many attempts have been made to analyze the individual pulses instead of the mixed temporal profiles. The most
efforts have been put on long GRBs with single pulses, double pulses or multi-pulses that are well distinguished and separated from each other or sGRBs
including just one single pulse.

Several fitting functional forms with different numbers of parameters are used in pulse-fitting procedures for
individual prompt emission and afterglows pulse
\citep{1996ApJ...459...393,2005ApJ...627...324,2000APJL...529...L13,2002APJ...566...210,2003APJ...596...389,2007APJ...662...1093,2007APJ...670...565}. They
explore many attributes of fitting-pulses including the FWHM, the pulse asymmetry, the spectral lag, the pulse amplitude, the peak time, etc., to classify long
and short bursts, or to study the prompt emission mechanisms. Wider pulses tend to be more asymmetrical \citep{1996ApJ...459...393}, while
\cite{2003APJ...596...389} found that there was no significant correlation between the asymmetry and the FWHM of pulses. The pulse amplitude is highly
anti-correlated with the other pulse timing features (e.g., the rise time of pulse, the fall time of pulse, the FWHM) \citep{2002AA...385...377}. For the long
BATSE bursts, the pulse width is positively correlated with both the peak time and the spectral lag \citep{2005ApJ...627...324}. \cite{2006MNRAS...373...729}
adopted a cross-correlation technique to measure the time lags of 65 single-peaked bursts and found that the lags of the vast majority of sGRBs were so small that
they were negligible or non-measurable. \cite{1995APJ...448...L101} assumed a power-law dependence of pulse width on energy. \cite{1996ApJ...459...393} drew the
similar conclusion with a sample of 41 bright long GRBs detected by BATSE. Other authors also studied the same power-law relation but they gave very different
indexes for their samples \citep{2006MNRAS...368...1351,2008APJ...685...1052,2015APJ...811...93,2017APJ...844...126}. In addition, some literatures had
investigated in details about the relation between the different timescales and temporal structures of GRBs pulses \citep{2004APJ...617...439,2007AN...328...99},
where three main timescales, namely
angular spreading timescale, cooling timescale and dynamic timescale, had been considered to form the shapes of GRB
pulses. They found that the curvature effect of the relativistic expanding spherical shell may play an important role in shaping the steep decay
phase while the dynamic timescale merely contributes to the rise phase in the observed pulses when the radiative time is
negligible \citep{2017APJ...840...95}. Therefore, the different timescales during the prompt $\gamma$-ray emission might be useful to diagnose the structure of
the jetted outflows according to the pulse evolution across diverse energy channels.

The GRB spectra and light-curves vary with the expansion of fireballs and the evolution of jet
structure \citep{2019MNRAS...482...5430,2002AA...396...705}. \cite{2019ApJ...876...89} found that there was a non-thermal pulse followed by a thermal component
for GRB 170817A. \cite{2017APJ...848...L14} noted that GRB 170817A had a lower peak energy than the average value of sGRBs and a near-median fluence.
\cite{2018APJ...869...55} constrained the outflow structure of GRB 170817A with a Markov Chain Monte Carlo Analysis and thought that a relativistic structured jet
with an opening angle of $\theta$$_0$ $\approx$5 degrees, a Lorentz factor of $\Gamma$ $\approx$175, and an off-axis angle of $\theta$$_{obs}$ = $27_{-3}^{+9}$
degrees, was largely favored. According to the multiple-wavelength observations of first gravitational-wave event GW170817/GRB 170817A, the cocoon model involving a
choked or structured jet cocoon has been widely accepted as the most likely jet mechanism after the neutron-star mergers from which sGRBs will be generally
produced \citep{2018Nature...554...207,2018Nature...561...355}.

In this paper, we give a comprehensive study of temporal attributes for the short GRB pulses from the fourth complete BATSE catalog. Particularly, we focus on the
structure analysis of the similarity or dissimilarity between single- or multiple-peaked sGRBs among different energy channels. Sample selection and data analysis
are presented in Section 2. Section 3 displays our main results. Some possibly physical explanations for the pulse evolution are given in Section 4. We will end
with conclusion and discussion in Section 5.
\section{DATA AND METHOD}
 Our initial sample consists of 100 sGRBs selected from the fourth BATSE TTE Catalog. The photon counts of these
 sGRBs have been accumulated within a time bin of 5 ms into four standard energy channels, that is Channel 1 (Ch1: 25-55keV), Channel 2 (Ch2: 55-110keV), Channel 3 (Ch3: 110-320keV), and Channel 4 (Ch4: $\geq$320keV). The selection criterion is that the peak count rate in any
 energy channels should be larger than 200 counts per second.

Based on many previous investigations, the pulses in long GRBs are typically asymmetric, i.e. fast rise and exponential decay (FRED). The pulse shapes of sGRBs
are vastly hard to determine precisely because of extreme short time variability, large range of signal to noise, and finite temporal resolution, in particular for some overlapping pulses. The GRB pulse functions
are however not uniquely defined. \cite{1996ApJ...459...393} proposed an asymmetric exponential rise and exponential decay function.
\begin{equation}\label{equation:1}
f(t)=\begin{cases}
 {Aexp[-(\frac{|t-t_{max}|}{\sigma_r})^\nu]}, t > t_{max}, \\
 {Aexp[-(\frac{|t-t_{max}|}{\sigma_d})^\nu]}, t < t_{max},
\end{cases}
\end{equation}
where \emph{A}, \emph{t}$_m$$_a$$_x$, $\sigma$$_r$, $\sigma$$_d$, $\nu$ are five parameters of a pulse, \emph{A} is the normalization parameter, \emph{t}$_{max}$
is the peak time, $\sigma$$_r$ and $\sigma$$_d$ are the rise and decay times, and $\nu$ measures the sharpness of the pulse. The pulse model had been easily applied to describe those spiky rather than smooth pulses of long GRBs mostly \citep [e.g.,][]{1999APJ...518...901,2014ApJ...783...88,2017APJ...844...126}.

From general experience of fitting purposes, \cite{2005ApJ...627...324} defined another form of pulse count rate that is proportional to the inverse of the
product of two exponentials, one increasing and one decreasing with time
\begin{equation}\label{equation:2}
\begin{split}
f(t)& ={A \lambda / [exp(\frac{\tau_1}{t})exp(\frac{t}{\tau_2})]} \\
& ={A \lambda exp(-\frac{\tau_1}{t}-\frac{t}{\tau_2}}), t > 0,
\end{split}
\end{equation}
where $\tau_1$ and $\tau_2$ are two modulating parameters connected with the rise and the decay steepness of a given pulse, $\lambda$=exp(2$\mu$) with
$\mu$=($\tau_1$/$\tau_2$)$^{1/2}$. At \emph{t}=$\tau$$_{peak}$=($\tau_1$$\tau_2$)$^{1/2}$, the intensity reaches its maximum that is normalized by $\lambda$ to
the peak intensity \emph{A}. The time of pulse onset with respect to \emph{t}=0 has been ignored. Previously, many authors had used this model to obtain a number of pulse properties (e.g., \citealt{Hakkila2008,2014ApJ...783...88,2015ApJ...815...134,2018ApJ...855...101}). Unfortunately, this model cannot be to employed to physically describe the GRB pulses whose asymmetries are characterized by longer rise than decay times \citep{2018ApJ...855...101}. It is easily to verify that the resulting pulses always exhibit a FRED-like profile at the $t>0$ region even though different peak times $\tau_{peak}$ have been assumed.

To describe the shape of long GRBs, \cite{2003APJ...596...389} proposed a more flexible pulse function (it was called the KRL model in
\cite{zhang2005}) with five parameters as,
\begin{equation}\label{equation:3}
\begin{split}
f(t)=f_m(\frac{t+t_0}{t_m+t_0})^r[\frac{d}{d+r}+\frac{r}{d+r}(\frac{t+t_0}{t_m+t_0})^{(r+1)}]^{-(\frac{r+d}{r+1})}\\
+f_0(t)
\end{split}
\end{equation}
in which $f_m$ represents the maximum flux of the GRB pulse and $t_m$ is the peak time. The quantities \emph{r} and \emph{d} are the two parameters describing
 the shape of an individual pulse. \emph{$t_0$} denotes the offset between the start of the first pulse and the trigger time, and $f_0(t)=at^2+bt+c$ represents a background
 at the observation time \emph{t}. By contrast, the KRL pulse model with six free parameters for constant background can reproduce any types of profiles \citep{2003APJ...596...389,zhang2005,2006MNRAS...373...729,2007AN...328...99,li2020} and is thus more flexible and universal if only the same real instead of ¡°incident¡± data from telescopes have been utilized.

Of course, which pulse fitting model is taken is largely irrelevant, given the obliging nature of GRB time profiles and the fact that the ``detected'' (processed through the detector response) rather ``incident'' flux is being modeled. Considering the above differences between three models and the variety of sGRB pulses, we empirically choose the KRL model for our current investigations. Due to the disturbance of white noises and backgrounds, the GRB pulses often become brighter with random fluctuations. So we adopt an adjacent-averaging method to smooth the original light curves and give further background substraction before fitting. Note that three points as the smoothing window have been chosen to extract the complete information of GRB pulses.

\begin{figure*}
\centering
\gridline{
\fig{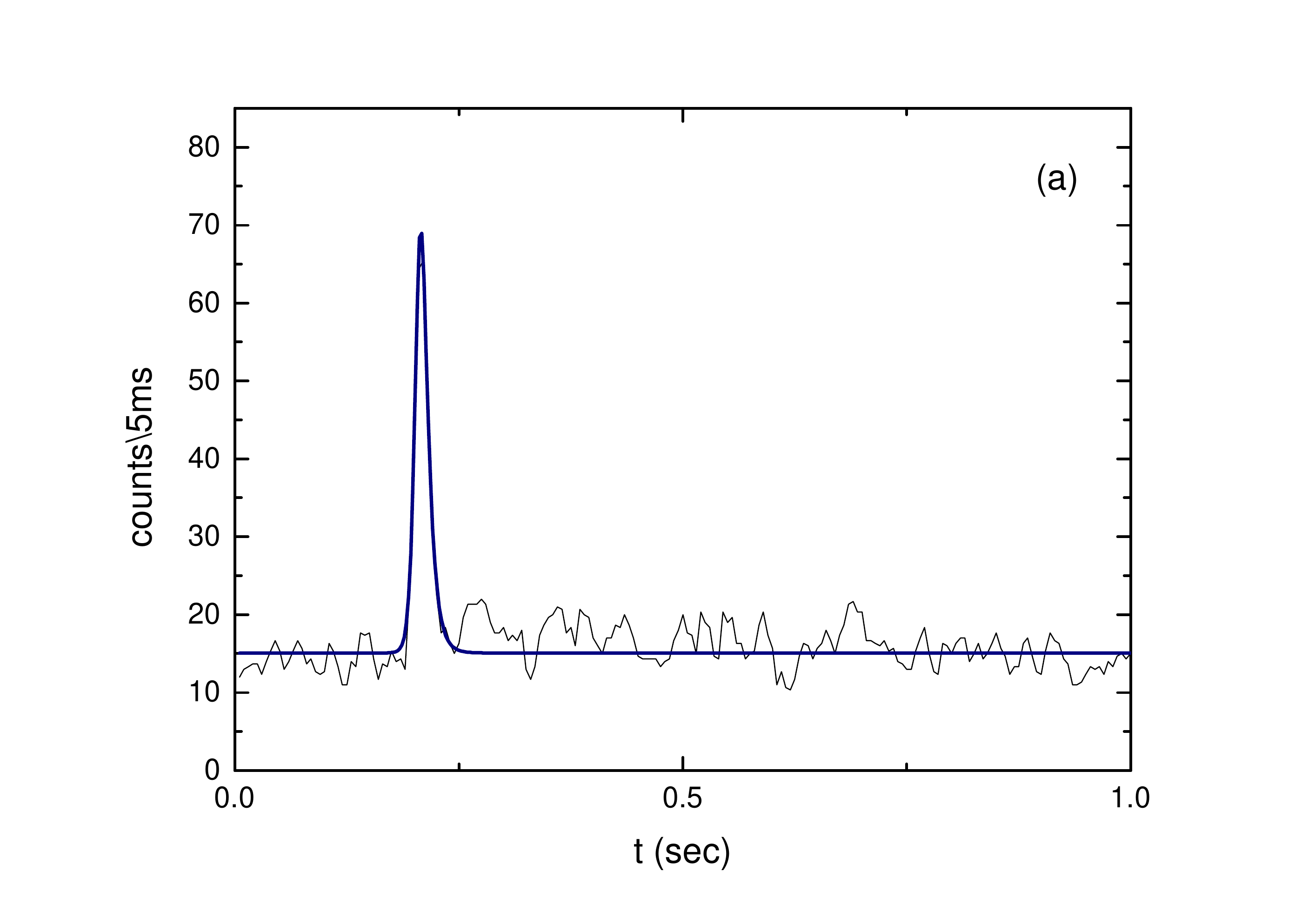}{0.4\textwidth}{}
\fig{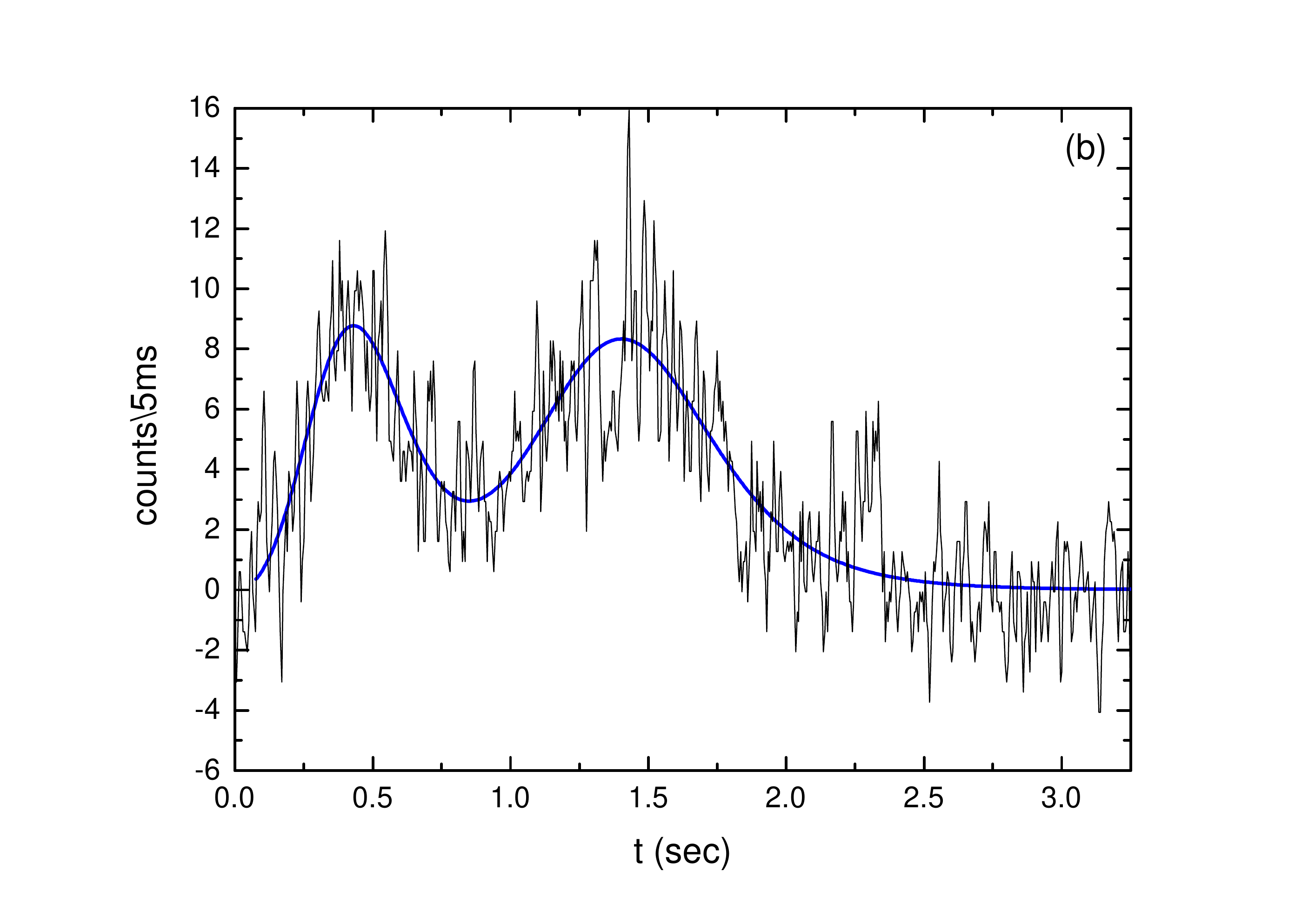}{0.4\textwidth}{}
          }
\gridline{
\fig{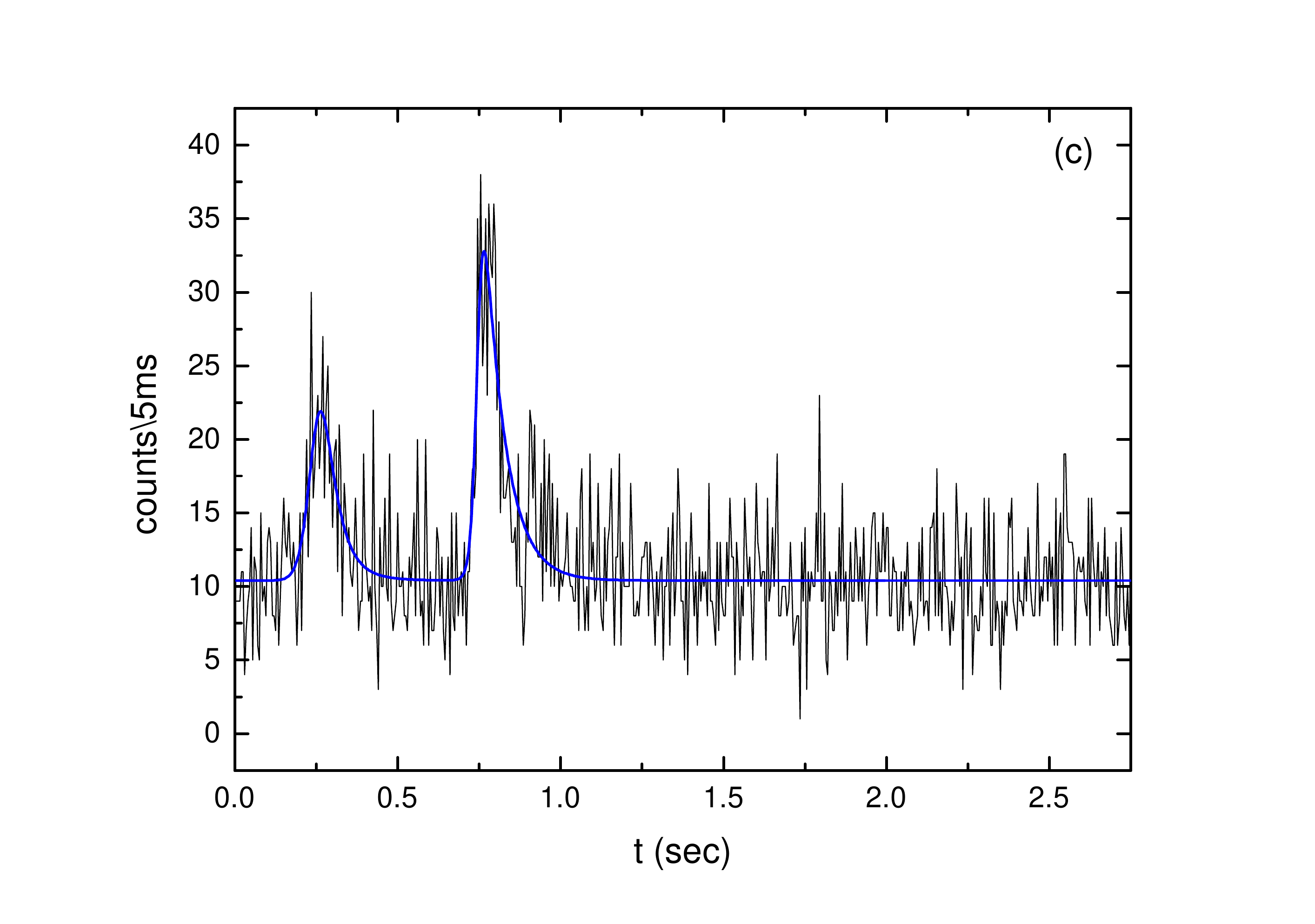}{0.4\textwidth}{}
\fig{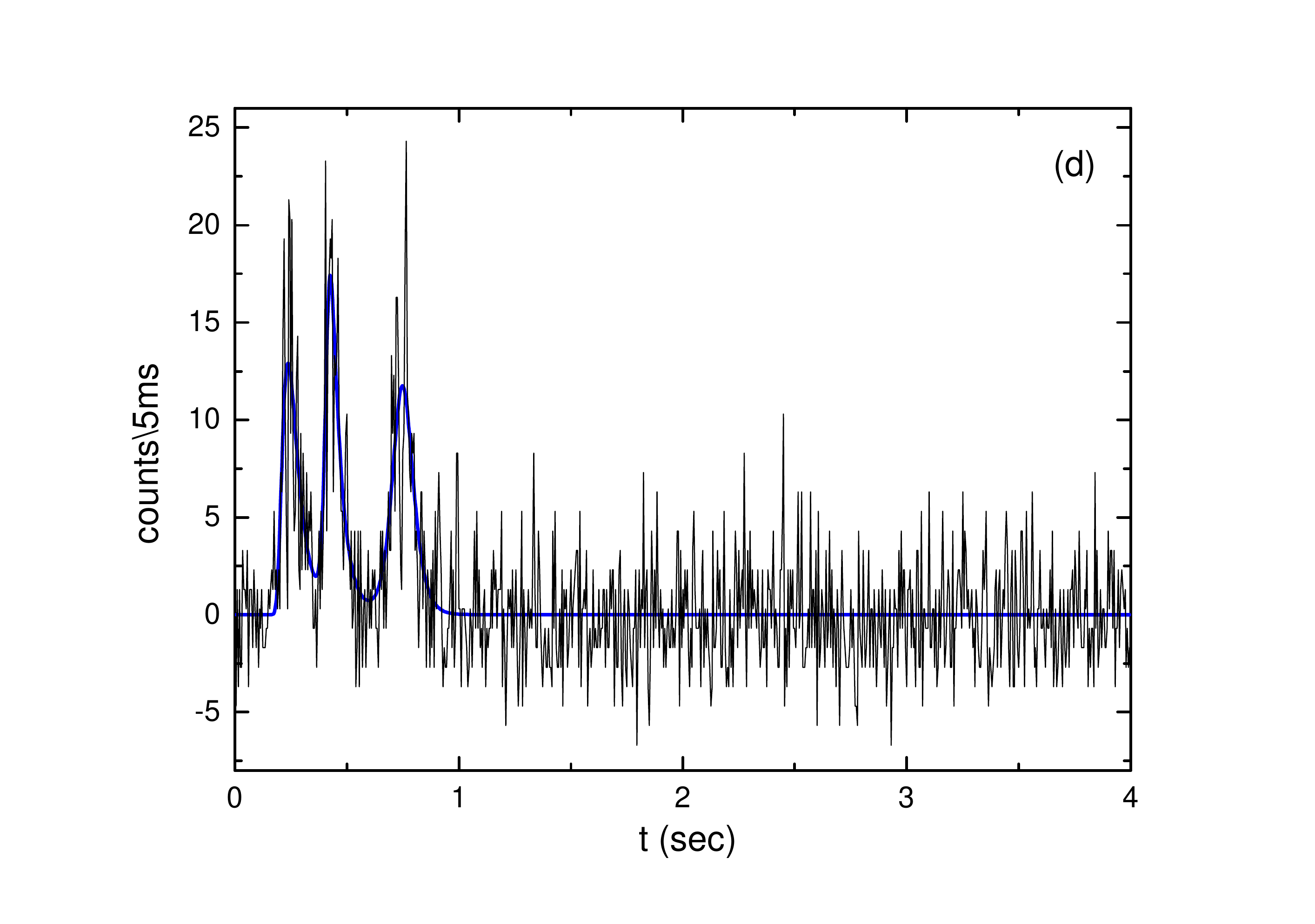}{0.4\textwidth}{}
                    }

\caption{Several typical light curves of the sGRBs. (a) SPs with trigger number $\#$ 0512 (Ch2). (b) Mt-DPs with trigger number $\#$ 03113 (Ch2). (c) Ml-DPs with
trigger number $\#$ 02115 (Ch3). (d) TPs with trigger number $\#$ 07102 (Ch3). \label{GRB0512-figure1}}
\end{figure*}

For this study, we firstly estimate the background level (1$\sigma$) and extract the effective pulse signals at a confidence level of $S/N>$ 3$\sigma$. Relatively, the single-peaked sGRBs called as SPs in Figure \ref{GRB0512-figure1} (a) are very easily distinguished. In the case of multiple-peaked sGRBs, we have used the least Chi-Square criterion to identify how many components within one burst as done by some previous authors, \citep [e.g.,][]{1996ApJ...459...393,2003APJ...596...389,2019ApJ...876...89}. Once a sGRB is distinguished to have double peaks (DPs) with a M-like shape, we then divide them into the M-tight types (Mt-DPs) in Figure \ref{GRB0512-figure1} (b) and the M-loose ones (Ml-DPs) shown in Figure \ref{GRB0512-figure1} (c), depending on whether the intensity percentage of trough of peaks is higher than 10\% or not (see also \citealt{2005ApJ...627...324}; \citealt{2019ApJ...876...89}). After excluding some sGRBs
with very complex structures, caused by the heavy overlapping between adjacent pulses, usually having more than three components, the refined sample with
well-fitting consists of 81 sGRBs including 57 SPs (57\%), 22 DPs (22\%), and 2 triple-peaked GRBs (TPs, 2\%) (see Figure \ref{GRB0512-figure1} (d)). Totally, 243
individual pulses have been carefully analyzed in details.

\section{Results}

\subsection{Pulse Properties}
We study the asymmetry, width, amplitude and energy dependence of sGRB pulses. For the SPs, pulses in Ch2 and Ch3 are compared to check the evolution of pulse
shape with the $\gamma$-ray energy. For two types of the DPs, the numbers of them are relatively limited so that we add together the channels 2 and 3 in order to
 increase statistical reliability.
\subsubsection{Asymmetry and Width }
\begin{figure*}
\centering
\gridline{
\fig{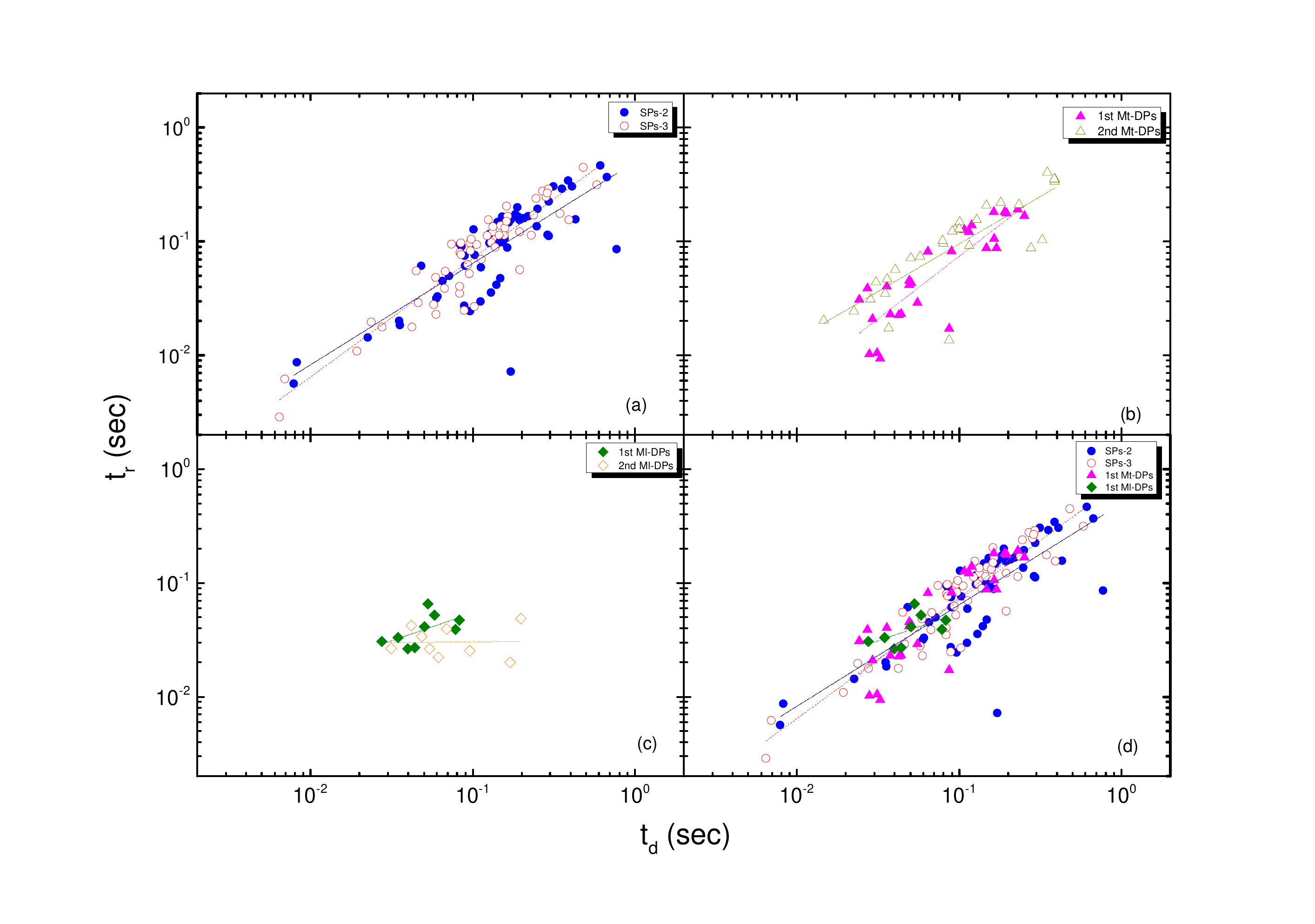}{0.6\textwidth}{}
          }
\caption{ Correlations between the t$_r$ and the t$_d$ for different kinds of sGRB pulses. (a) The filled and the open circles represent the
SPs in Ch2 and Ch3, respectively. (b) The filled and the open triangles represent the 1st and the 2nd pulses of the Mt-DPs. (c) The filled and the open diamonds
represent the 1st and the 2nd pulses of the Ml-DPs. (d) Comparisons between the SPs and the 1st pulses of both the Ml-DPs and the Mt-DPs. All the lines stand for
the best fits to the data.
\label{trtd-figure2}}
\end{figure*}

As usual, the asymmetry of a pulse is defined to be a ratio of the rise time $t_r$ of full width at half maximum (FWHM) to the decay time $t_d$
\citep[e.g.,][]{2001AA...380...L31,2007APSS...310...19}. It is noticeable that with the asymmetry is difficult enough to determine in lGRBs \citep{2003APJ...596...389}, and can be much more so in sGRBs. The systematic errors of pulse measurements are estimated with error propagation according to \cite{2006ChJAA...6...312}. Figure \ref{trtd-figure2} shows that the $t_r$ and the $t_d$ are good in agreement with a power-law
relation of t$_r \sim t_d$$^\beta$ for the SPs and the DPs, except the Ml-DPs. The power-law indexes are listed in Table \ref{table1}. Interestingly, these
results for sGRBs are much similar to those of long GRBs \citep{1996ApJ...459...393}. For the Ml-DPs, the power law correlation is less tight than that of the
Mt-DPs, especially for the second pulses whose t$_r \sim t_d$$^\beta$ seems to be inexistent. It needs to emphasize that the power law correlation between the
$t_r$ and the $t_d$ of the first pulses in the DPs is very consistent with that of the SPs. In addition, the power law correlations in both the first and the
second pulses of the Mt-DPs are relatively tighter than those of the Ml-DPs. It is also found that the measured parameters of DPs have relatively large statistical uncertainties than those of the SPs, hinting a trend of more pulse numbers more difficult to measure. Meanwhile, we find that the critical parameters $r$ and $d$ of sGRB pulses obviously hold larger scatters in contrast with lGRB pulses.

\begin{figure*}
\centering
\gridline{
\fig{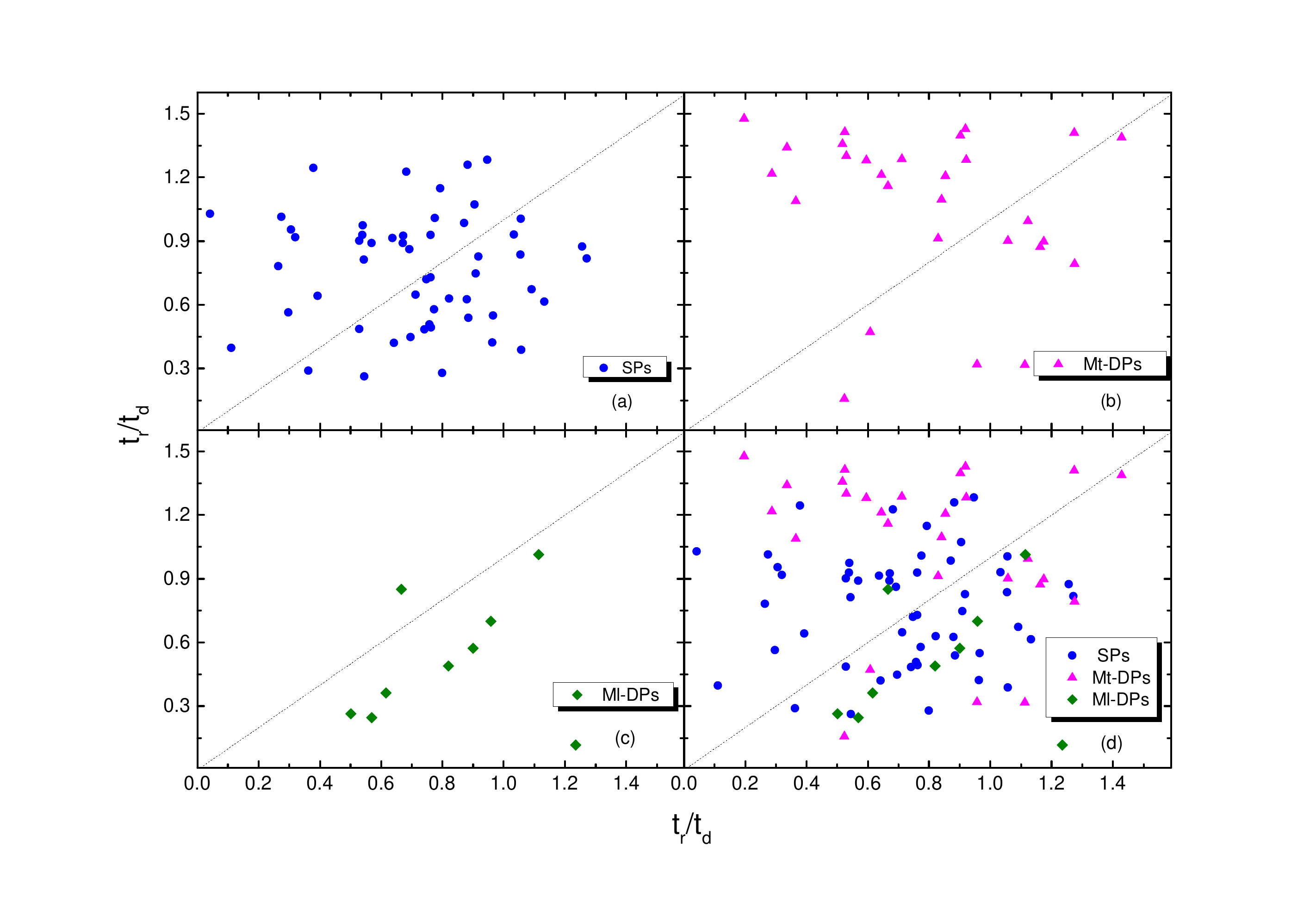}{0.6\textwidth}{}
          }
\caption{Comparisons of the t$_r$/t$_d$ for all three kinds of sGRB pulses. (a) asymmetries of the SPs pulses in Ch2 versus
Ch3. (b) asymmetries of the 1st versus the 2nd pulses in the Mt-DPs. (c) asymmetries of the 1st versus the 2nd pulses in the Ml-DPs. (d)  comparison of the
asymmetries between the SPs, the Ml-DPs and the Mt-DPs.
\label{asymmetry-figure3}}
\end{figure*}

From Figure \ref{asymmetry-figure3} and Table \ref{table2}, we obtain that the asymmetries of all three kinds of sGRBs range from 0.04 to
1.48 and have a mean (median) value of 0.79 (0.81). This result is slightly larger than the value of 0.65 measured for a sample of 100 bright BATSE sGRBs by
\cite{2001AA...380...L31}, however, it is quite close to the value of 0.81 given for the single-peaked sGRBs \citep{2007APSS...310...19}. Especially, we find the
mean asymmetry of the SPs is very similar with that of the
1st pulses of the two kinds of DPs. It is worthy to point that the averaged asymmetries of the SPs and the 2nd pulses in the DPs are largely different, which
implies that the SPs and the 1st pulses in DPs could share the same origin. For the two TPs in trigger $\#$06715 and $\#$07102, their pulse asymmetries will
evolve with time, which hints that the second or third pulse could be produced from different emitting regions.

\begin{figure*}
\centering
\gridline{
\fig{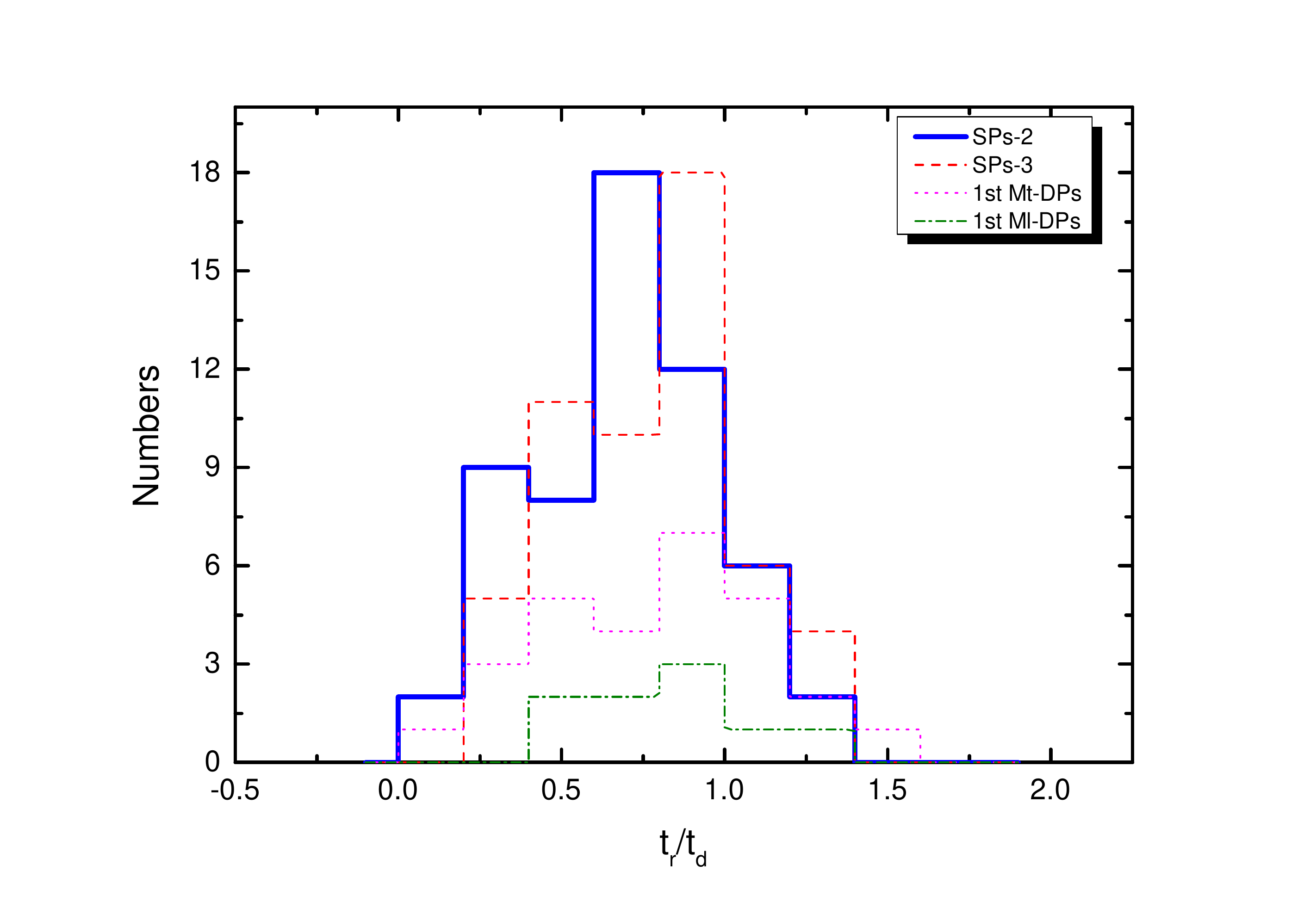}{0.6\textwidth}{}
          }
\caption{Distributions of the t$_r$/t$_d$ of all three kinds of sGRB pulses: the SPs pulses in Ch2 (solid line) and Ch3 (dash line), the 1st pulses of the Mt-DPs
(dot line) and the 1st pulses of the Ml-DPs (dash dot line)
\label{disasymmetry-figure4}}
\end{figure*}

\begin{figure*}
\centering
\gridline{
\fig{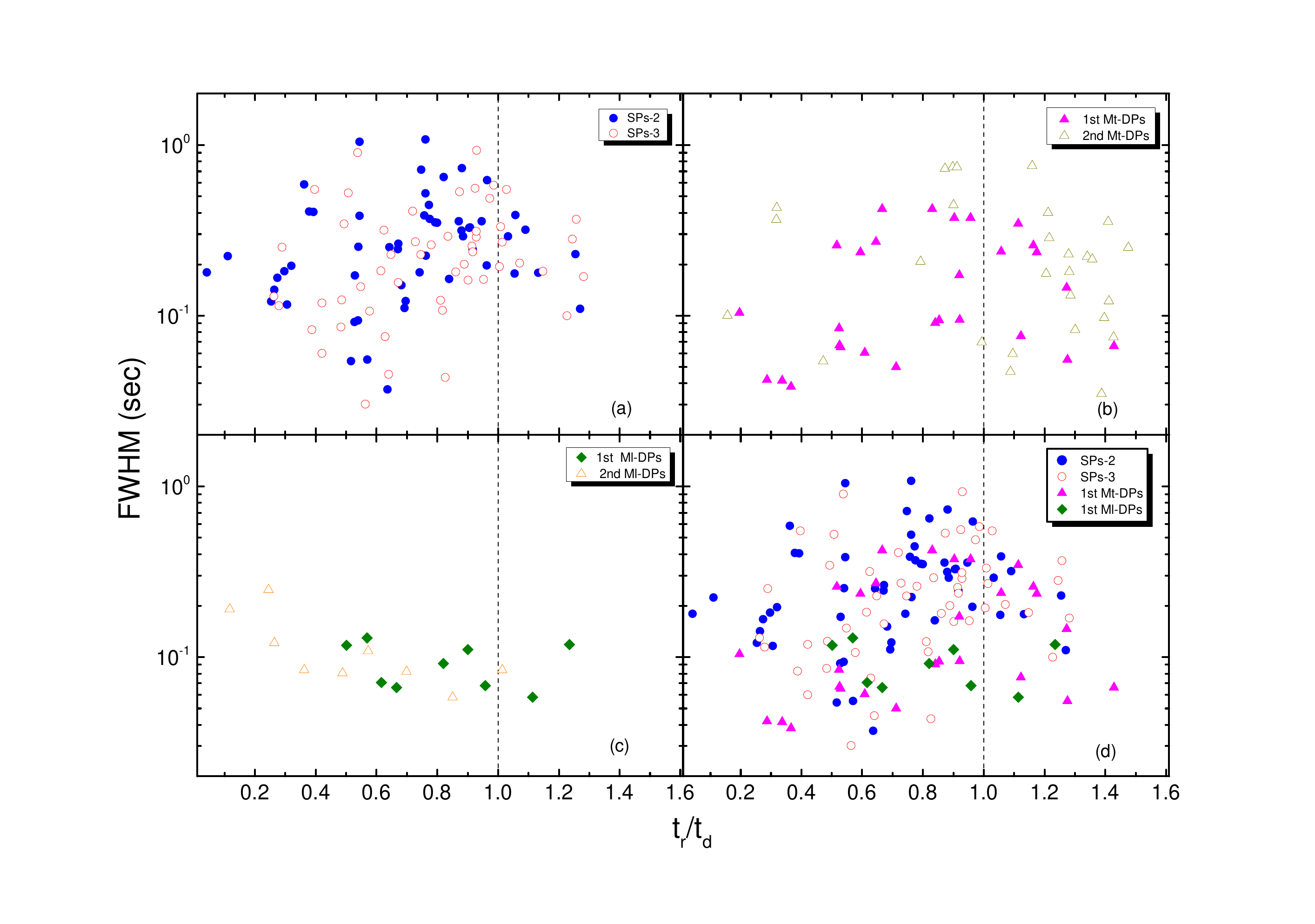}{0.6\textwidth}{}
          }
\caption{Correlations between the FWHM and the t$_r$/t$_d$ of all three kinds of sGRB pulses. The symbols are the same as Figure 2.
\label{FWHMasy-figure5}}
\end{figure*}

Figure \ref{disasymmetry-figure4} shows the distributions of t$_r/t_d$ for the SPs pulses and the 1st pulses of the two
subclasses of the DPs. A K-S test gives the probabilities of P$_{SP2,SP3} \sim $ 0.37 with D$_1 \sim $ 0.17, P$_{SP2, Mt1} \sim $ 0.37 with D$_2 \sim $ 0.20,
P$_{SP2, Ml1} \sim $ 0.84 with D$_3 \sim $ 0.20, indicating that the four distributions are drawn from the same parent distribution. Figure \ref{FWHMasy-figure5}
shows that the pulse width and the
asymmetry are not correlated for the sGRB pulses. This phenomenon is similar to that found by \cite{2003APJ...596...389}, while it is inconsistent with
\cite{1996ApJ...459...393} for long bursts. Surprisingly, both pulses in the Ml-type sGRBs, unlike those in the Mt-DPs and the SPs, exhibit a coincident weak
anti-correlation between the FWHM and the asymmetry, that is more wide more asymmetric, which may hint a distinct geometrical or physical mechanism for these
kinds of bursts. Note that the dependence of the FWHM on the asymmetry in the second pulse is relatively weaker than that in the first pulse of the Ml-DPs.

\begin{figure*}
\centering
\gridline{
\fig{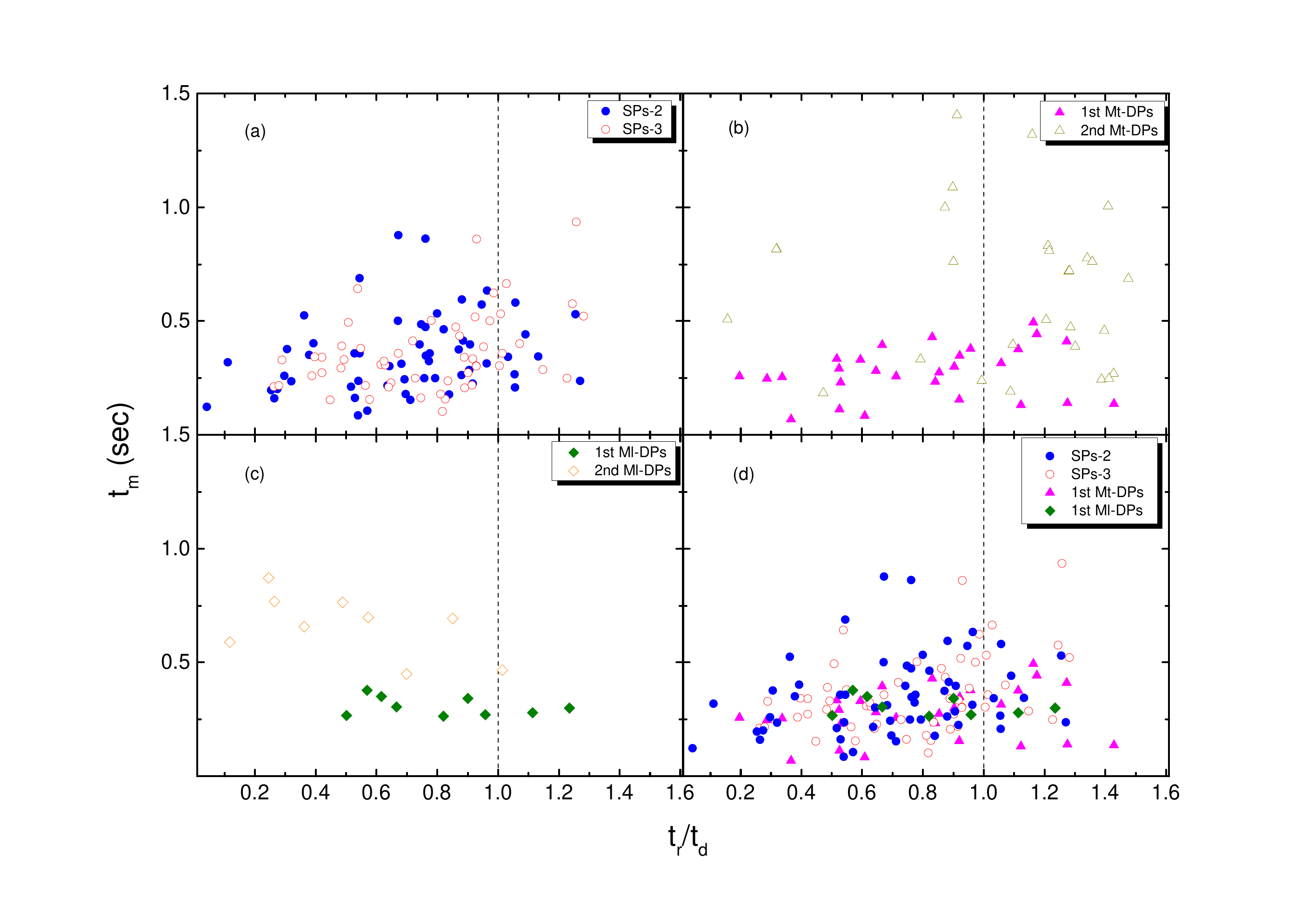}{0.6\textwidth}{}
          }
\caption{Correlations between the t$_m$ and the t$_r$/t$_d$ of all three kinds of sGRB pulses. The symbols are the same as Figure 2.
\label{tmasy-figure6}}
\end{figure*}

Moreover, we study the relation of the peak time (t$_m$) with the asymmetry appears to also have no significant law in Figure \ref{tmasy-figure6}, from which we
notice that the t$_m$ and the t$_r/t_d$ are independent for the SPs, the Mt-DPs and the 1st pulses in the Ml-DPs. It is surprisingly found the t$_m$ is
anti-correlated with t$_r/t_d$ for the 2nd pulse in the Ml-DPs, as shown in Figure \ref{FWHMasy-figure5} Panel (c) in Figures.
~\ref{FWHMasy-figure5}$-$~\ref{tmasy-figure6} seems to show that the 2nd pulses in the Ml-DPs have a diverse formation mechanism from other types of pulses.

\begin{figure*}
\centering
\gridline{
\fig{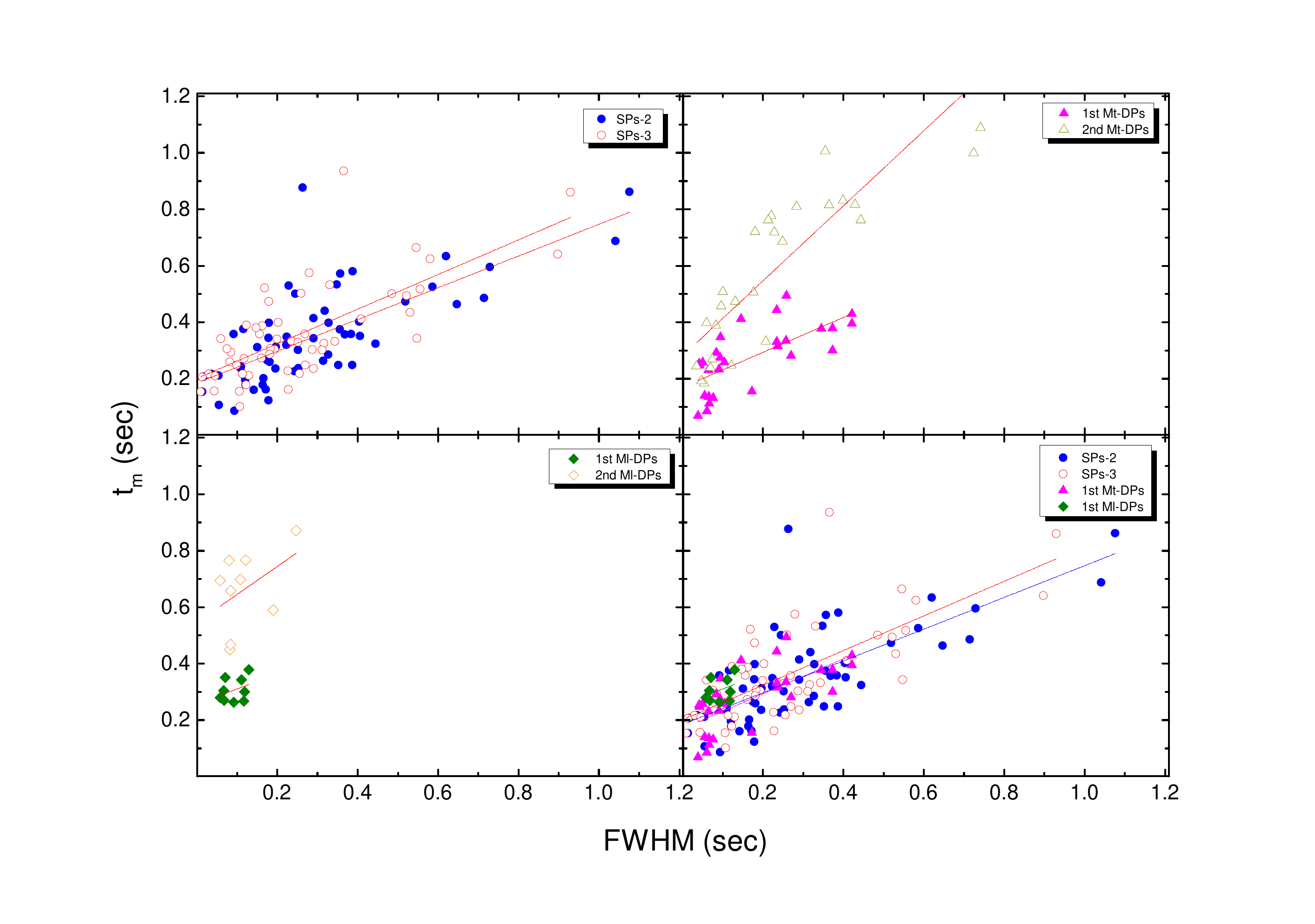}{0.6\textwidth}{}
          }
\caption{Correlations between the t$_m$ and the FWHM of all three kinds of sGRB pulses. The symbols are as same as Figure 2. The lines are the best fits.
\label{tmFWHM-figure7}}
\end{figure*}

Figure \ref{tmFWHM-figure7} indicates that the t$_m$ and the FWHM are positively correlated with a form as t$_m \sim \mu  FWHM$, which is consistent
with the previous result of lGRBs of BATSE \citep{2005ApJ...627...324}. The fitting results are listed in Table \ref{table3}. It proves again that the
characteristics of the 1st pulse of two subclasses of the DPs and the SPs are very similar and hence they may share the same origin from the central engine.

\subsubsection{Pulse Amplitude}
\begin{figure*}
\centering
\gridline{
\fig{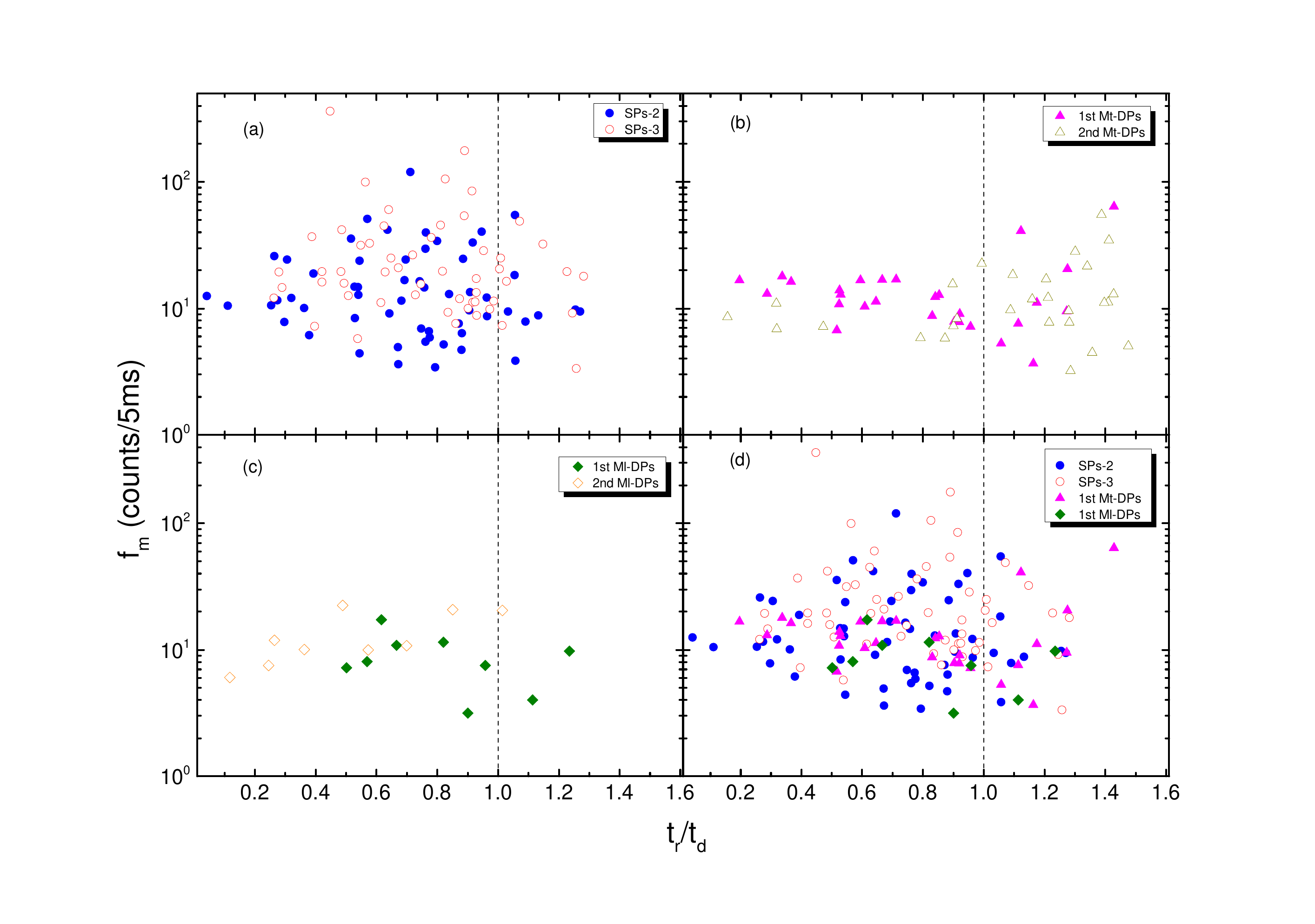}{0.6\textwidth}{}
          }
\caption{Correlations between the pulse amplitude f$_m$ and the t$_r$/t$_d$ of all three kinds of sGRB pulses. The symbols are the same as Figure 2.
\label{fmasy-figure8}}
\end{figure*}

\begin{figure*}
\centering
\gridline{
\fig{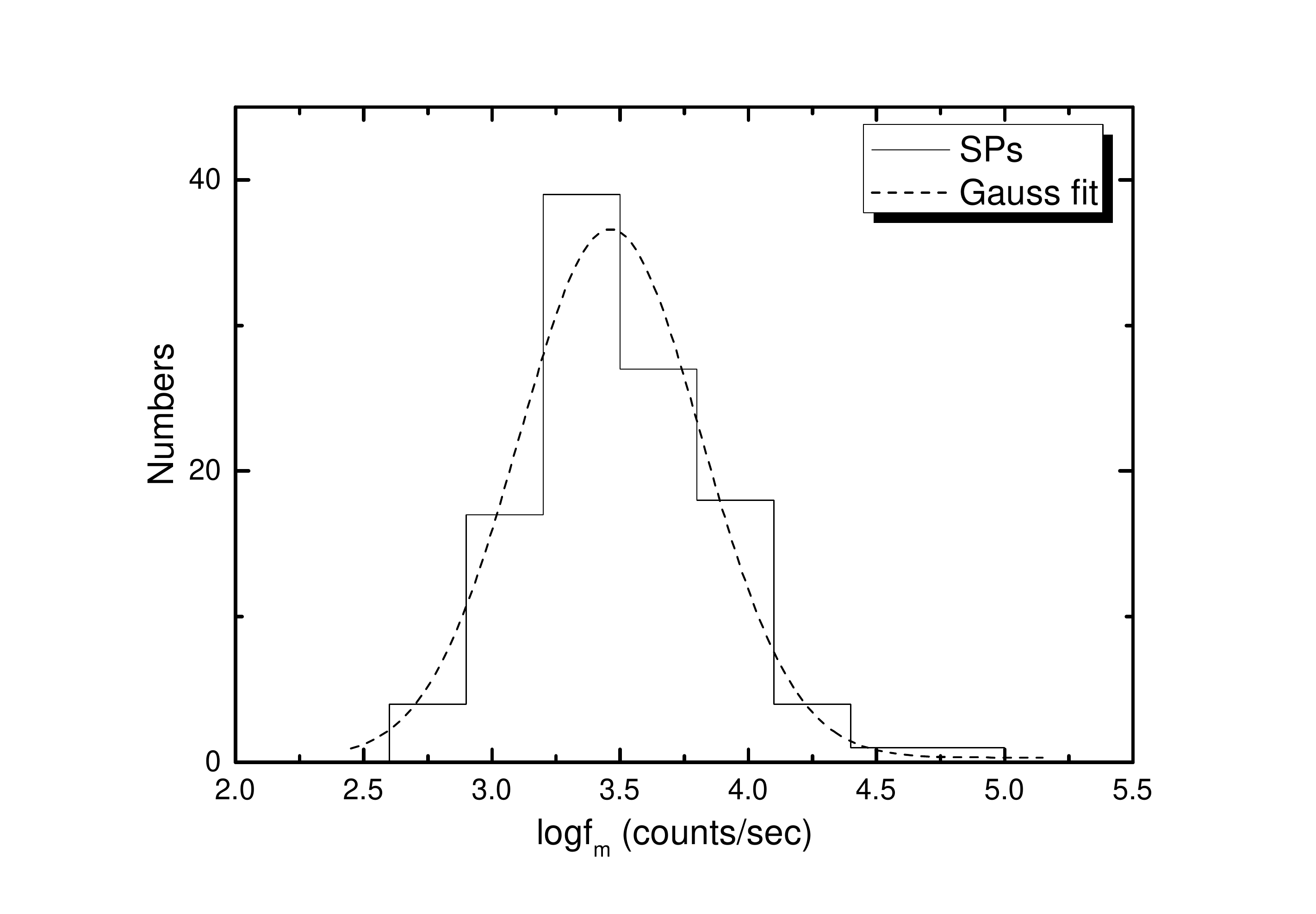}{0.6\textwidth}{}
          }
\caption{Distribution of the f$_m$ of the 111 SPs. The dash line represents the best fit with a log-normal function.
\label{singlelogfm-figure9}}
\end{figure*}

\begin{figure*}
\centering
\gridline{
\fig{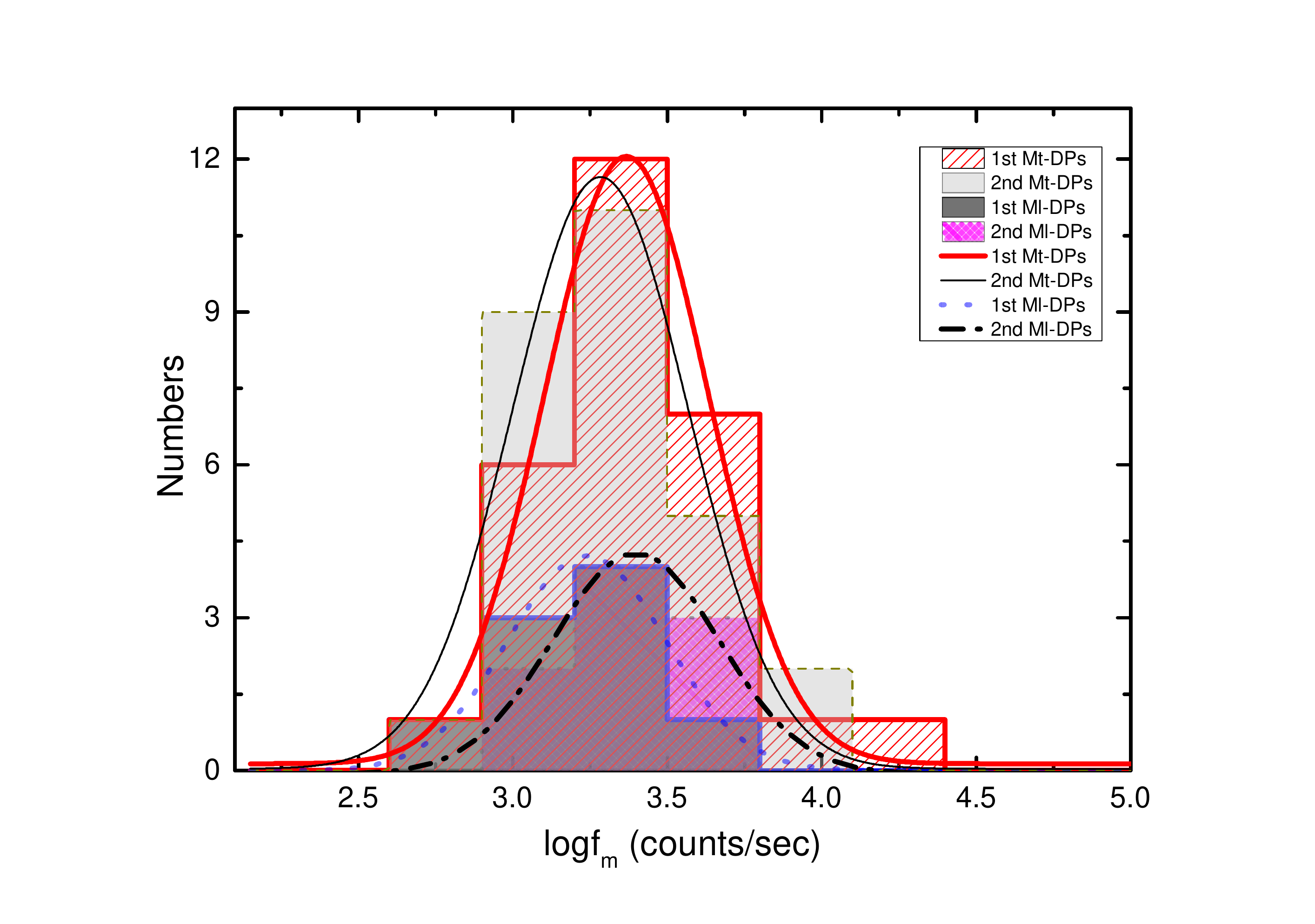}{0.6\textwidth}{}
          }
\caption{Distributions of the logf$_m$ for the 9 Ml-DPs and the 28 Mt-DPs. On the upper region, the light gray and hatched areas correspondingly represent the
distributions of the 1st and the 2nd pulses of the Mt-DPs; On the lower region, the dark gray and dense hatched areas represent the distributions of the 1st and
the 2nd pulses of the Ml-DPs, respectively. The solid and dash lines are the best fits with a log-normal function to the distributions.
\label{logfmdouble-figure10}}
\end{figure*}

\begin{figure*}
\centering
\gridline{
\fig{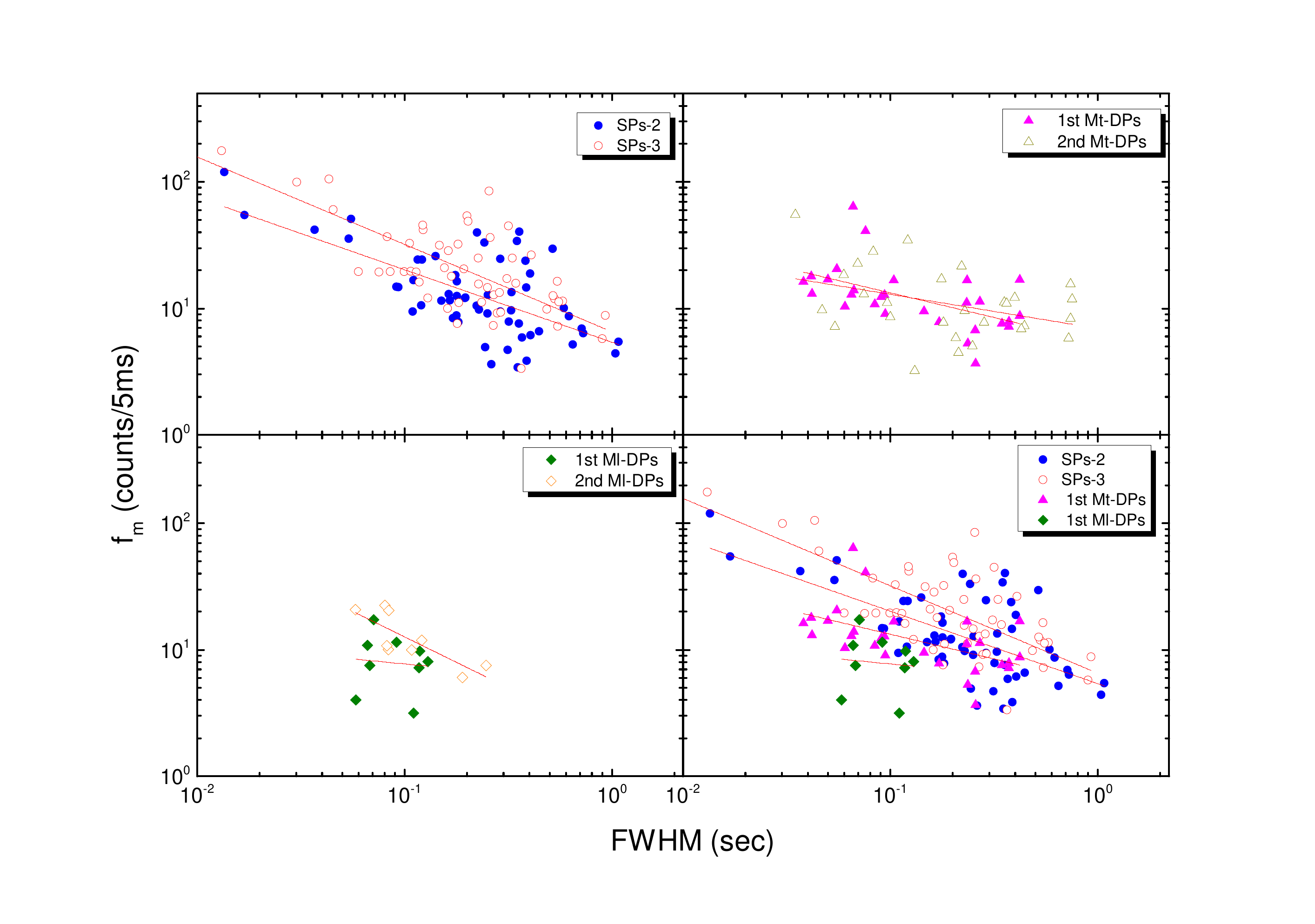}{0.6\textwidth}{}
          }
\caption{Correlations between the f$_m$ and the FWHM of all three kinds of sGRB pulses. The symbols are the same as Figure 2.
\label{fmFWHM-figure11}}
\end{figure*}

The pulse amplitude (f$_m$) reflects the released energy amount of the interactions between internal shocks. Figure \ref{fmasy-figure8} demonstrates that there is
no obvious correlations between the f$_m$ and the t$_r/t_d$ for either the SPs or the DPs. Figure \ref{singlelogfm-figure9} indicates that the f$_m$ of the SPs is
lognormally distributed with a mean value of f$_{m,SP}$=2884.03$^{+206.27}_{-192.50}$ counts/s. Figure \ref{logfmdouble-figure10} gives the mean values of the
f$_m$ to be f$_{m,Mt1}$=2344.23$^{+54.60}_{-53.36}$ counts/s for the 1st pulses and f$_{m,Mt2}$=1905.46$^{+44.80}_{-85.76}$ counts/s for the 2nd pulses of the
Mt-DPs. For the Ml-DPs, the mean values of the f$_m$ are f$_{m,Ml1}$=1737.80$^{+40.48}_{-39.56}$ counts/s in the 1st pulses and
f$_{m,Ml2}$=2511.89$^{+118.38}_{-113.06}$ counts/s in the 2nd pulses. The results indicate that the released energy amount in the 1st pulses of the Mt-DPs is
usually larger than that in the 2nd pulses while it is opposite  for the Ml-DPs, which demonstrates that these two kinds of DPs might originate from different
physical processes in essence.  Figure \ref{fmFWHM-figure11} shows that the f$_m$ and the FWHM are anti-correlated with a power-law form of $f_m \sim FWHM^\nu$
that is consistent with the previous conclusion drawn for long GRBs \citep{2002AA...385...377}. The power-law indexes are listed in Table \ref{table4}. For the
Ml-DPs, the power law index of the 1st pulse is larger than that of the 2nd one. On the
contrary, for the Mt-DPs, the index of the 1st pulses is smaller than that of the 2nd one. The opposite result suggests again
that the two M-type DPs might be produced by different kinds of mechanisms originally.

\subsection{Relation of width with Energy}
To test whether the power law relation, $FWHM\propto E^{\alpha}$, of the pulse width with the photon energy hold for the sGRBs, we select those qualified sGRBs
detected in three channels at least to ensure the relation between the FWHM and the photon energy can be successfully constructed. Since the photon fluxes in Ch1
and Ch4 are usually too weak to be detected significantly, we sometimes need to combine the two channels with Ch2 and Ch3, respectively. Considering the errors of
\emph{f$_m$, t$_m$, r} and \emph{d}, the error of the FWHM is estimated with error propagation according to our previous work \citep{2006MNRAS...373...729}. In
general, we adopt the averaged photon energy in the first three channels and 500
keV to be the estimated energy of photons in the fourth channel. Table \ref{table5} lists our detailed fitting results.

\begin{figure*}
\centering
\vspace{-0.3cm}
\subfigure{
\begin{minipage}{4.5cm}
\centering
\includegraphics[width=5cm]{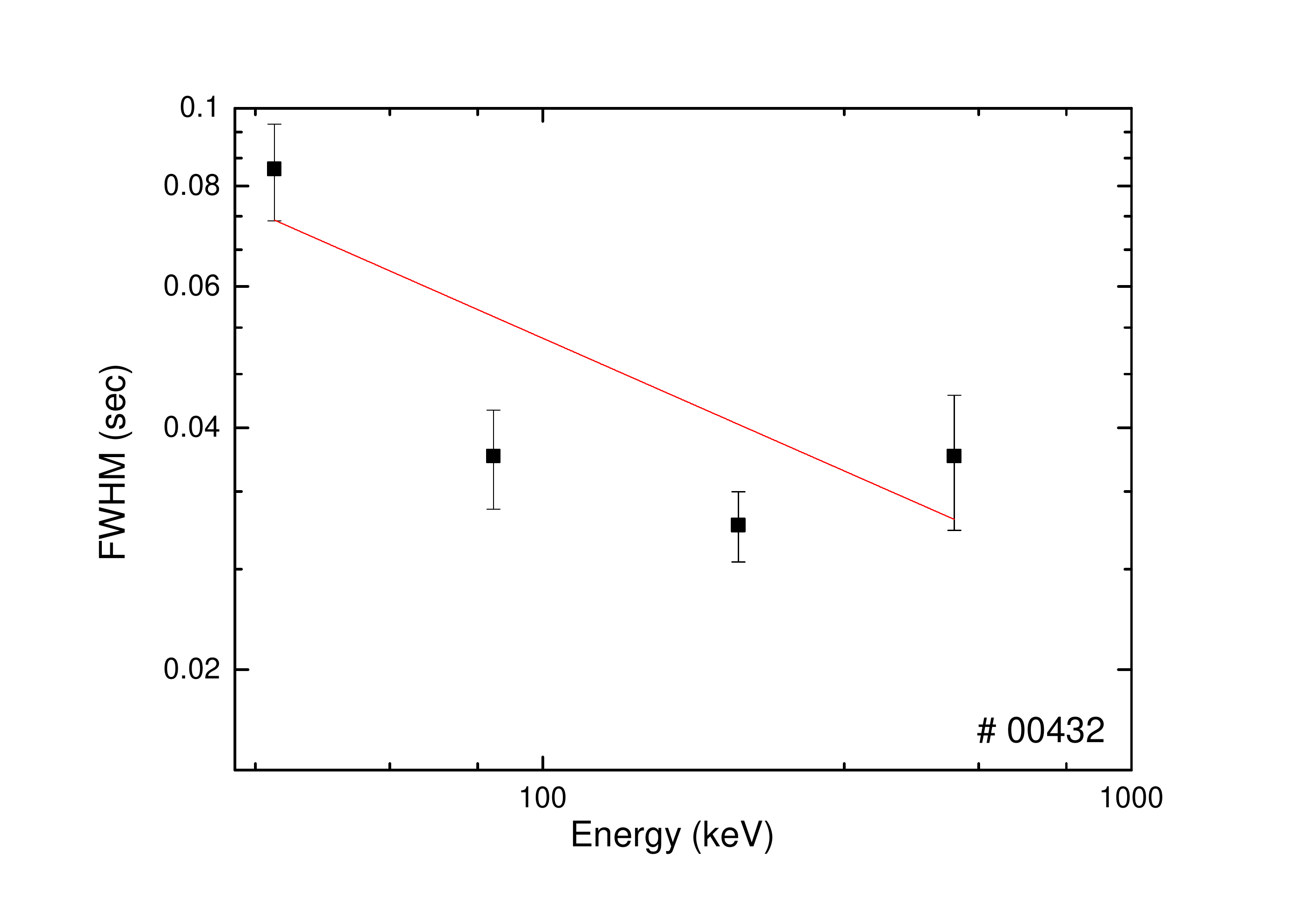}
\end{minipage}
}
\vspace{-0.3cm}
\subfigure{
\begin{minipage}{4.5cm}
\centering
\includegraphics[width=5cm]{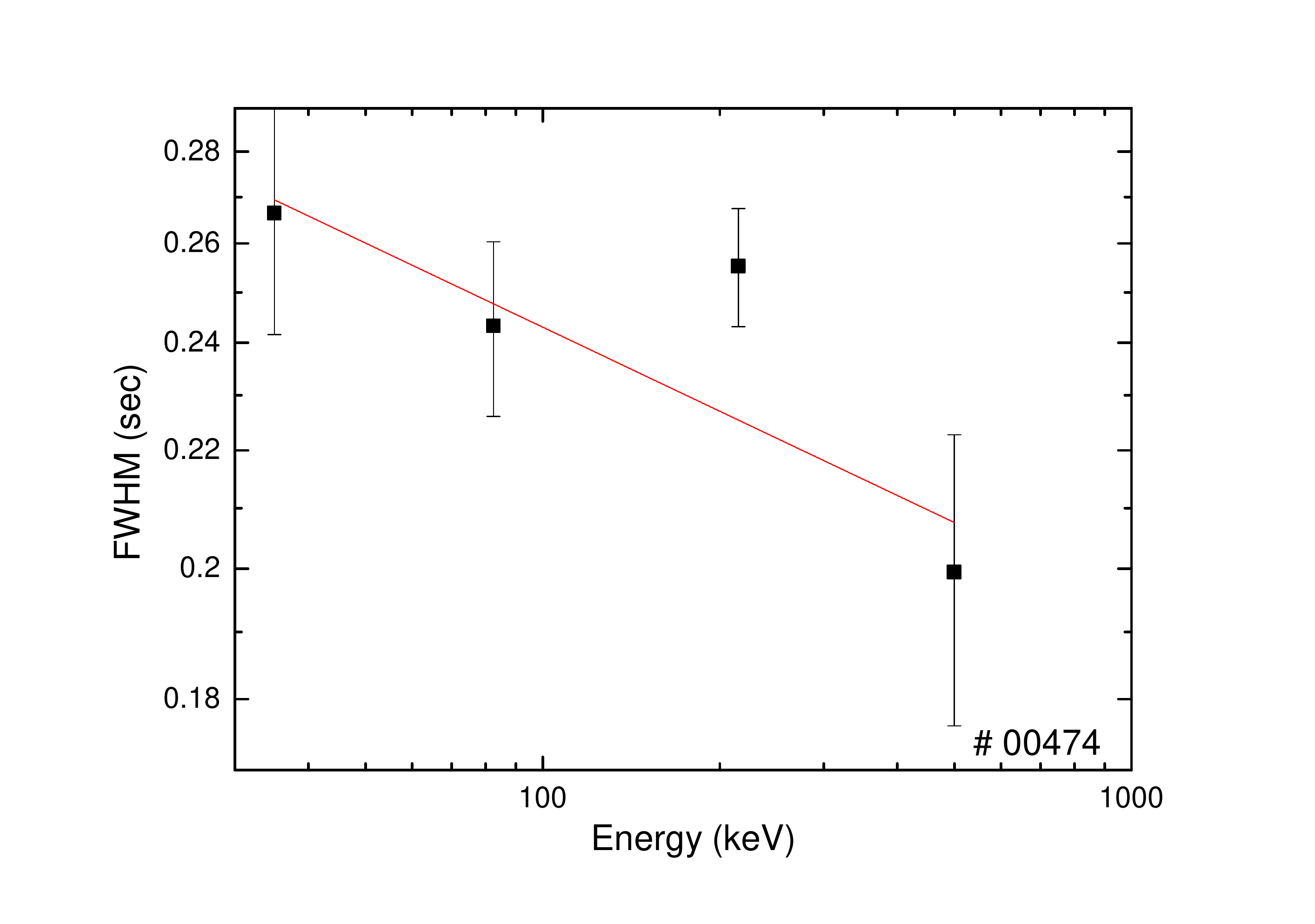}
\end{minipage}
}
\vspace{-0.1cm}
\subfigure{
\begin{minipage}{4.5cm}
\centering
\includegraphics[width=5cm]{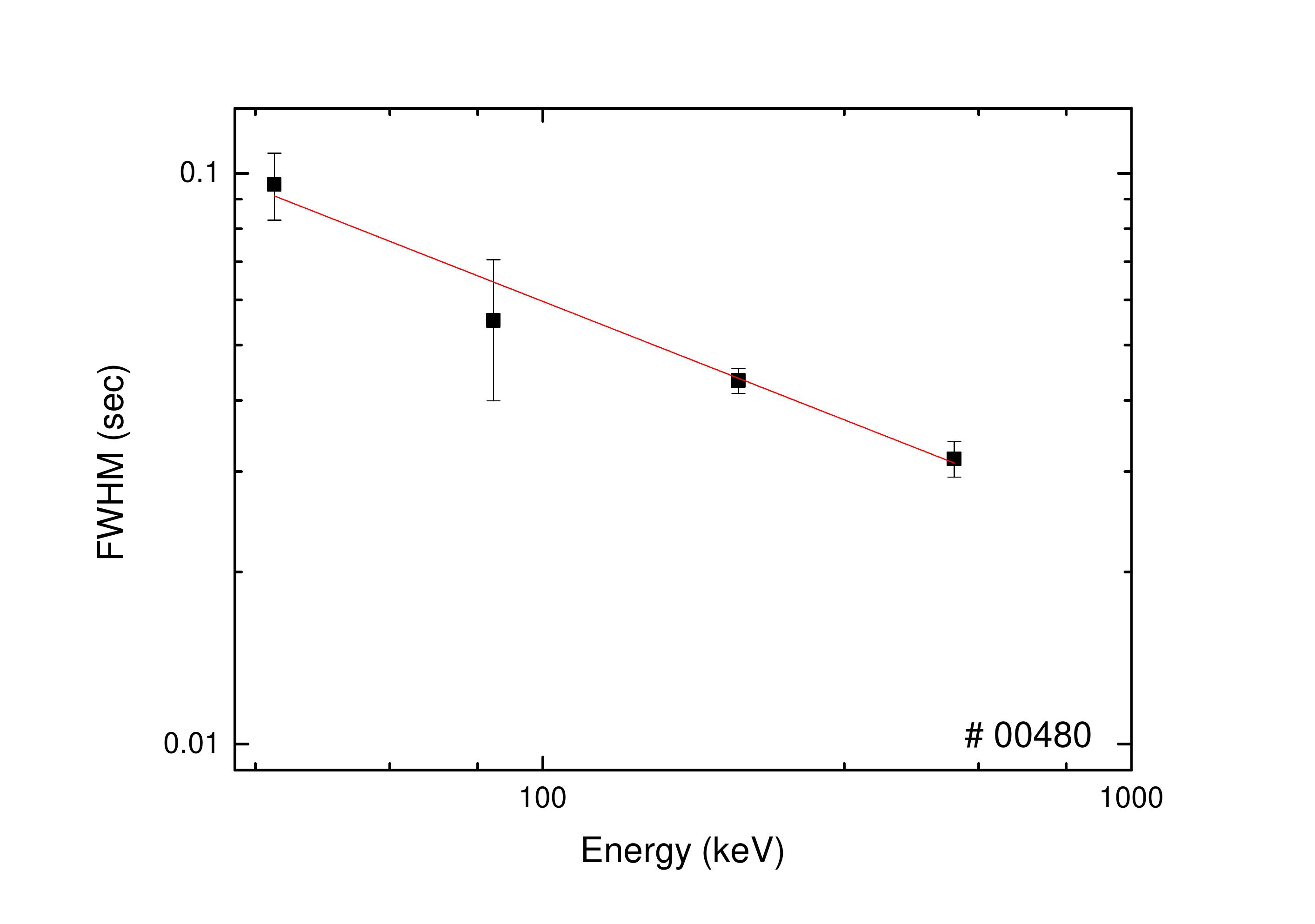}
\end{minipage}
}
\vspace{-0.1cm}
\subfigure{
\begin{minipage}{4.5cm}
\centering
\includegraphics[width=5cm]{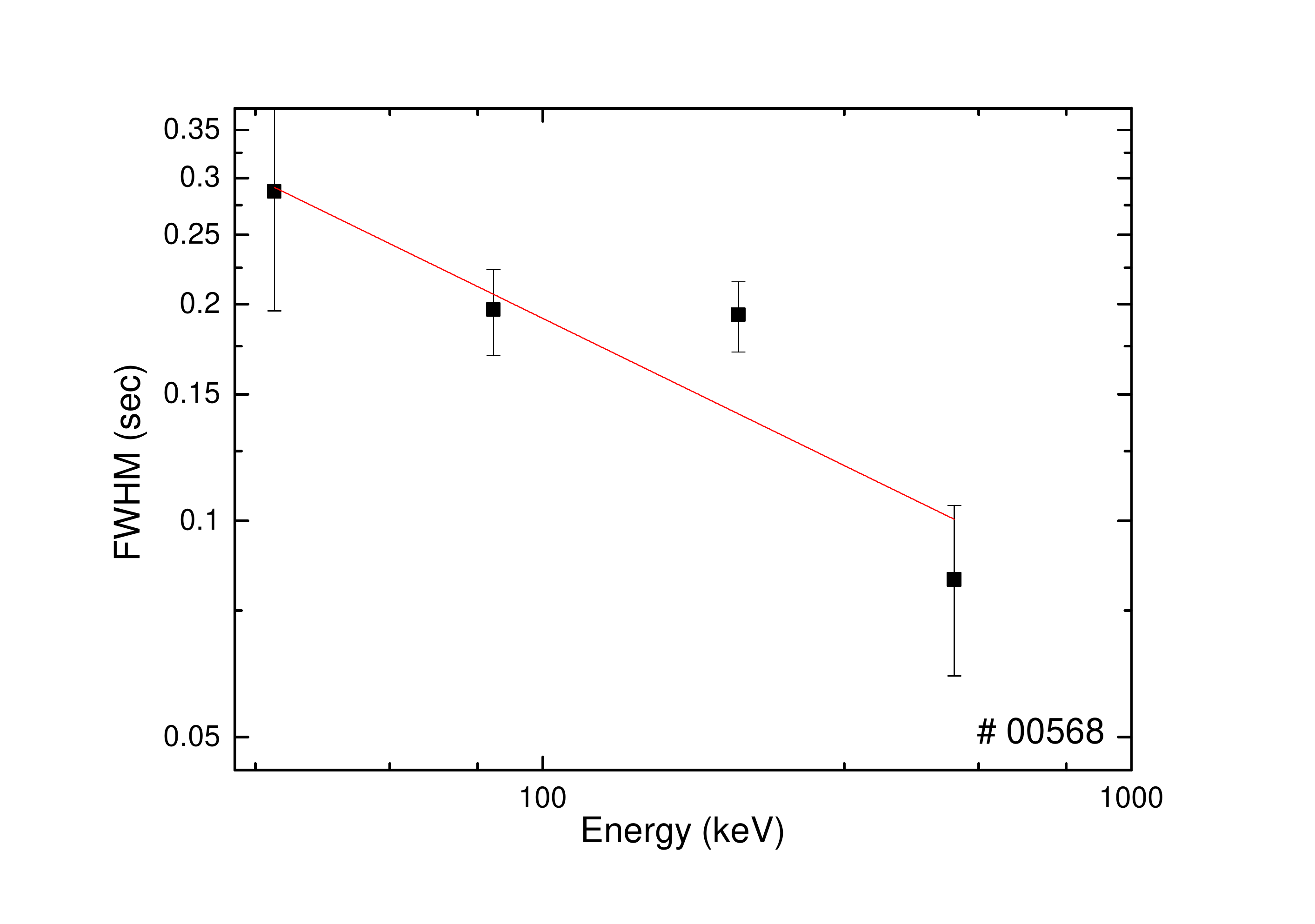}
\end{minipage}
}
\vspace{-0.1cm}
\subfigure{
\begin{minipage}{4.5cm}
\centering
\includegraphics[width=5cm]{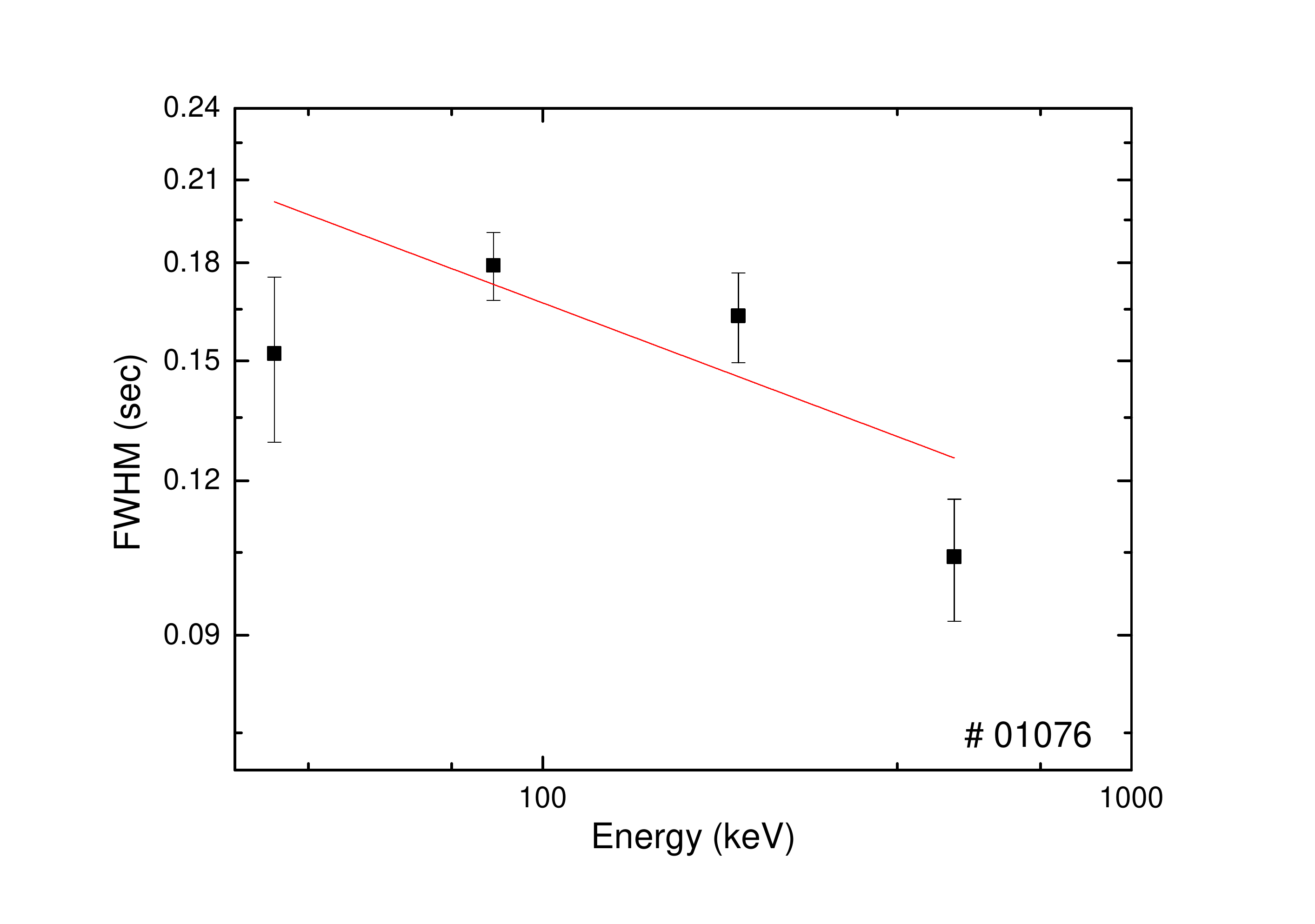}
\end{minipage}
}
\vspace{-0.1cm}
\subfigure{
\begin{minipage}{4.5cm}
\centering
\includegraphics[width=5cm]{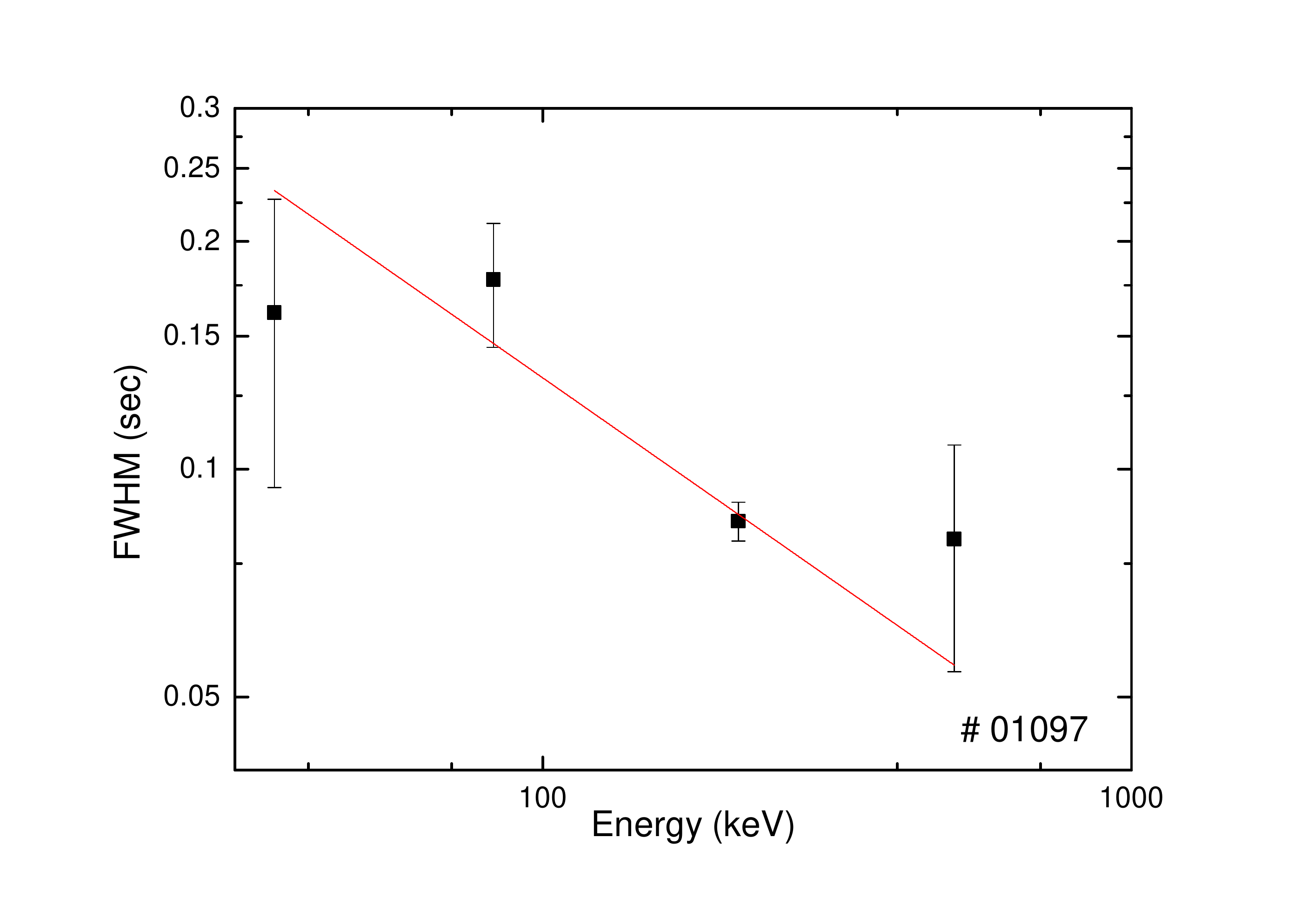}
\end{minipage}
}
\vspace{-0.1cm}
\subfigure{
\begin{minipage}{4.5cm}
\centering
\includegraphics[width=5cm]{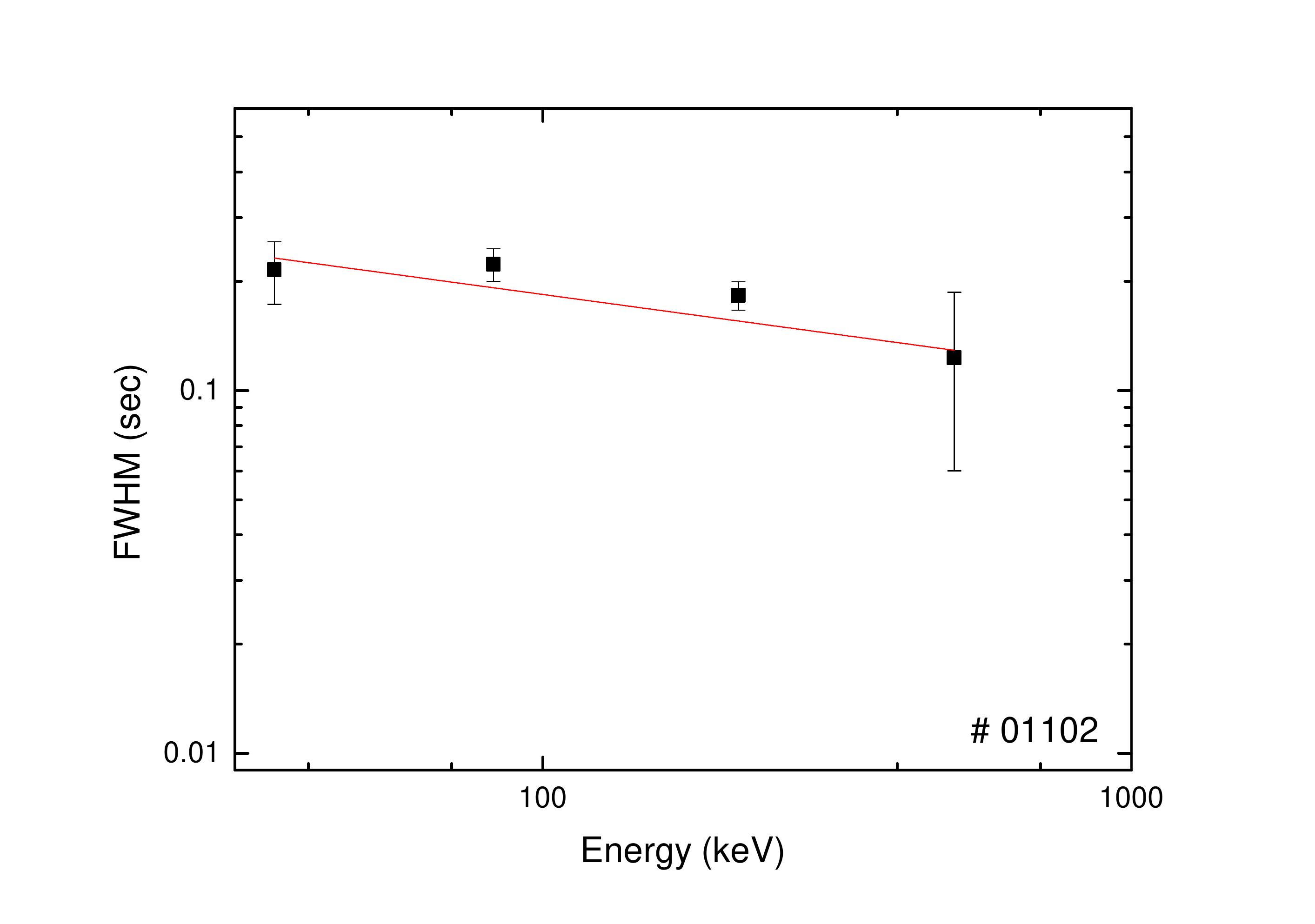}
\end{minipage}
}
\vspace{-0.1cm}
\subfigure{
\begin{minipage}{4.5cm}
\centering
\includegraphics[width=5cm]{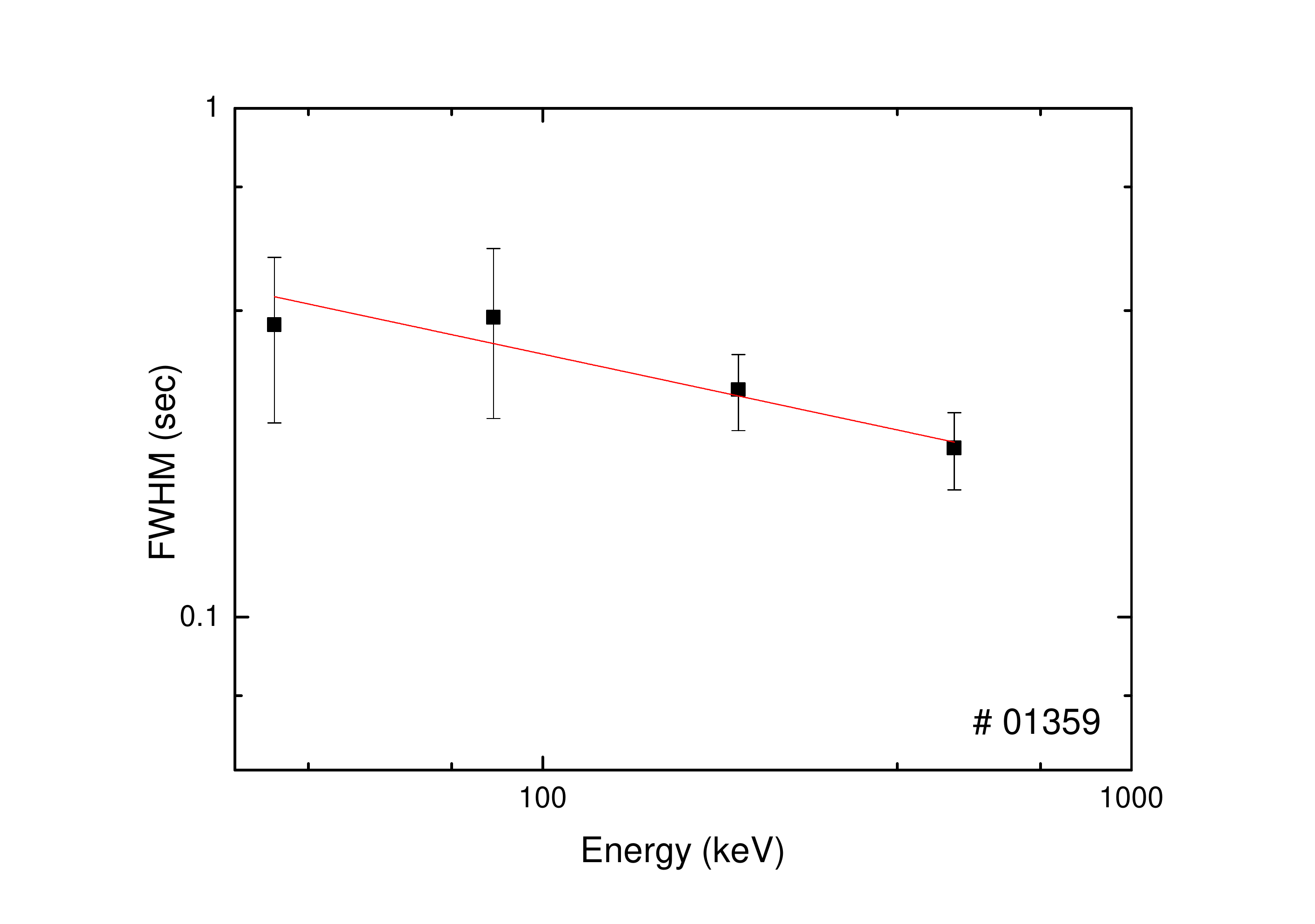}
\end{minipage}
}
\vspace{-0.1cm}
\subfigure{
\begin{minipage}{4.5cm}
\centering
\includegraphics[width=5cm]{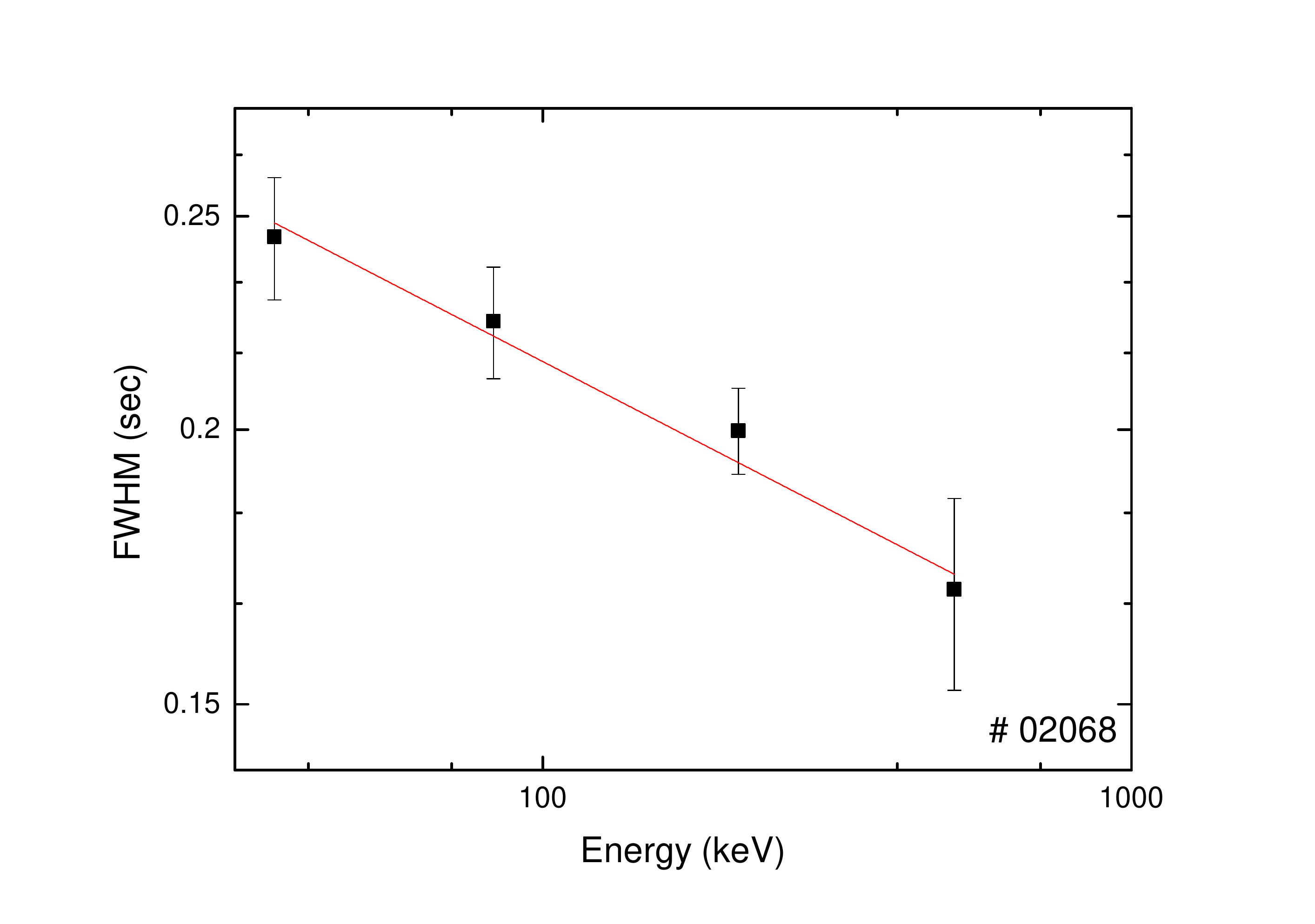}
\end{minipage}
}
\vspace{-0.1cm}
\subfigure{
\begin{minipage}{4.5cm}
\centering
\includegraphics[width=5cm]{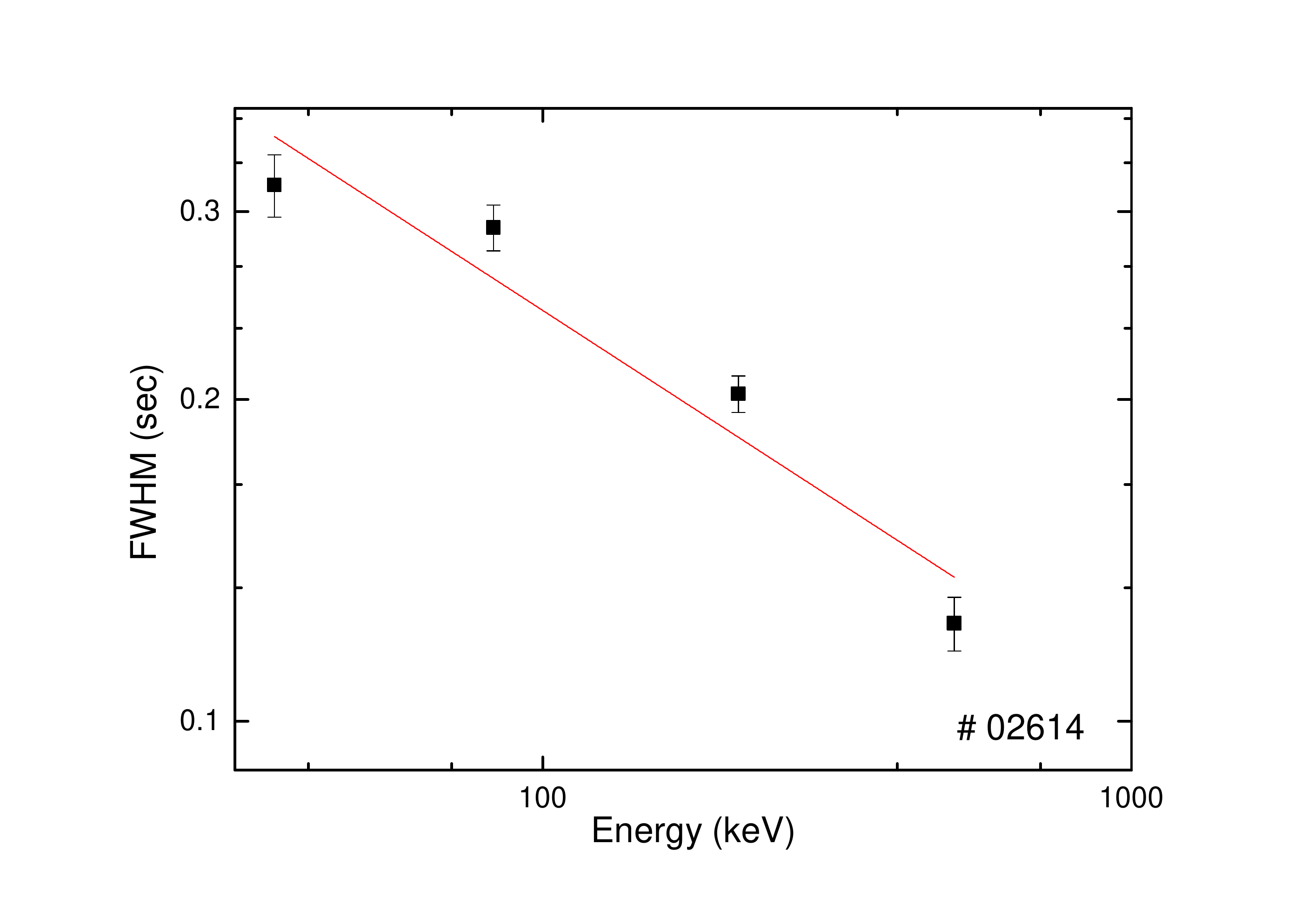}
\end{minipage}
}
\vspace{-0.3cm}
\subfigure{
\begin{minipage}{4.5cm}
\centering
\includegraphics[width=5cm]{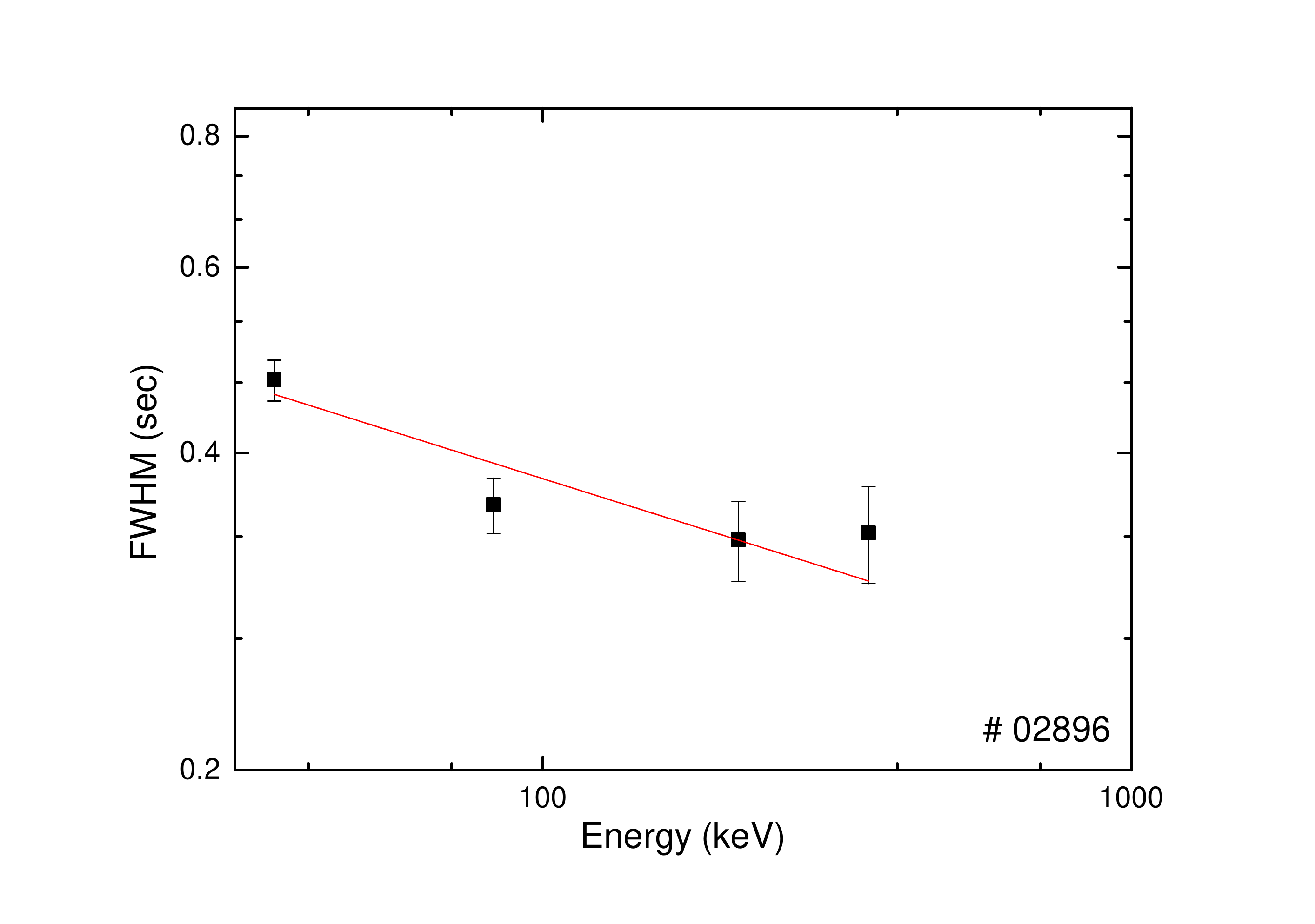}
\end{minipage}
}
\vspace{-0.1cm}
\subfigure{
\begin{minipage}{4.5cm}
\centering
\includegraphics[width=5cm]{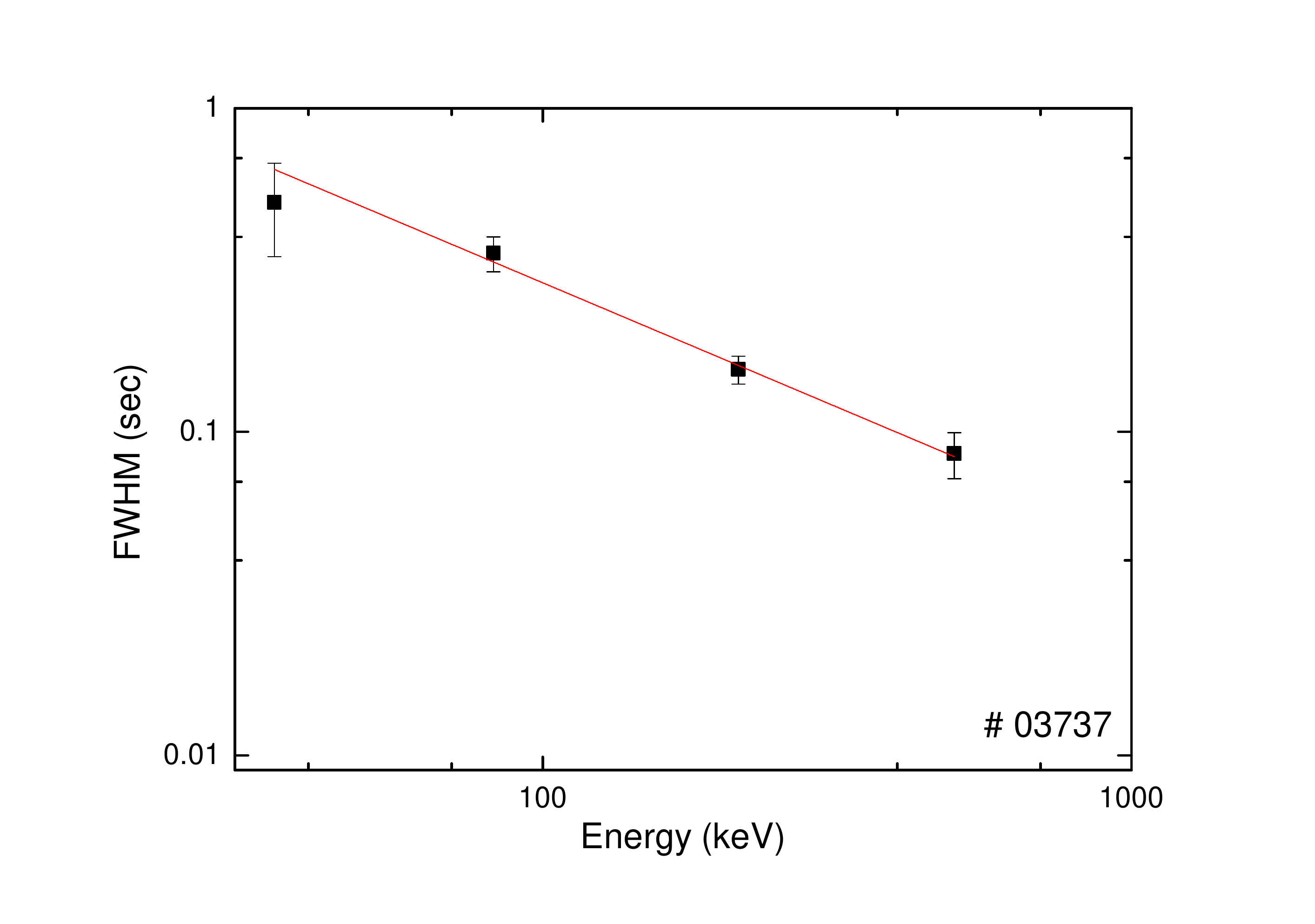}
\end{minipage}
}
\vspace{-0.1cm}
\subfigure{
\begin{minipage}{4.5cm}
\centering
\includegraphics[width=5cm]{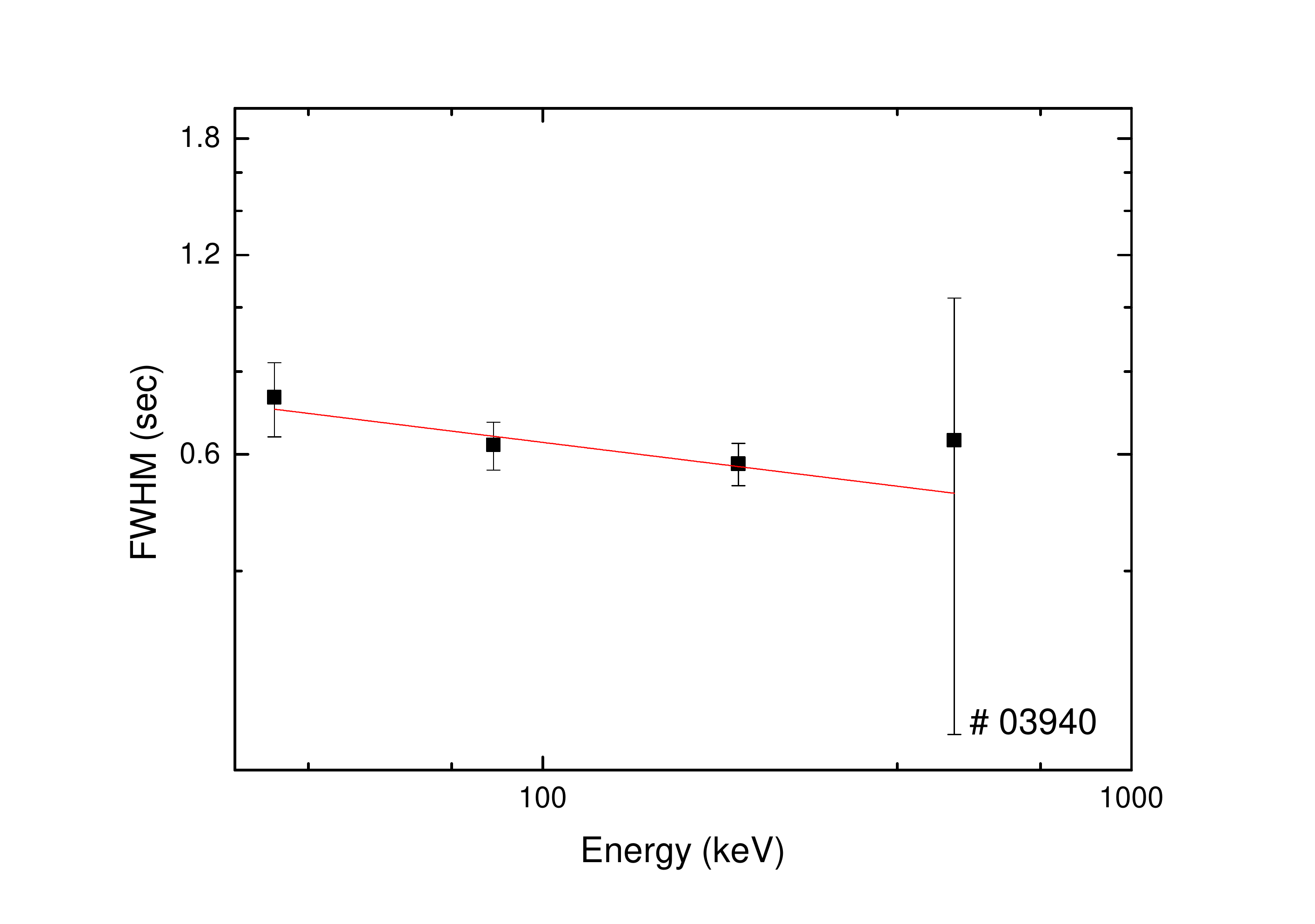}
\end{minipage}
}
\vspace{-0.1cm}
\subfigure{
\begin{minipage}{4.5cm}
\centering
\includegraphics[width=5cm]{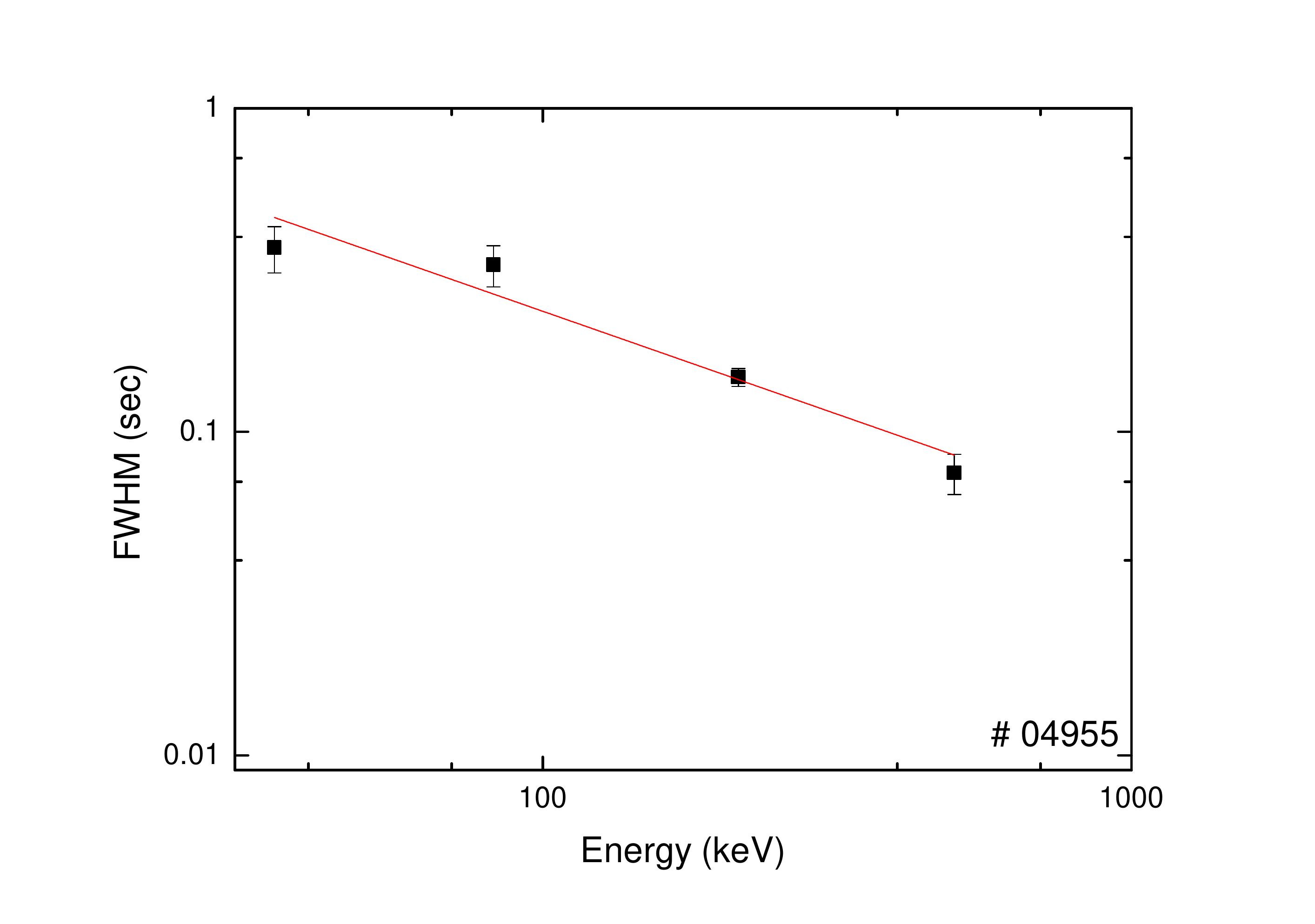}
\end{minipage}
}
\vspace{-0.1cm}
\subfigure{
\begin{minipage}{4.5cm}
\centering
\includegraphics[width=5cm]{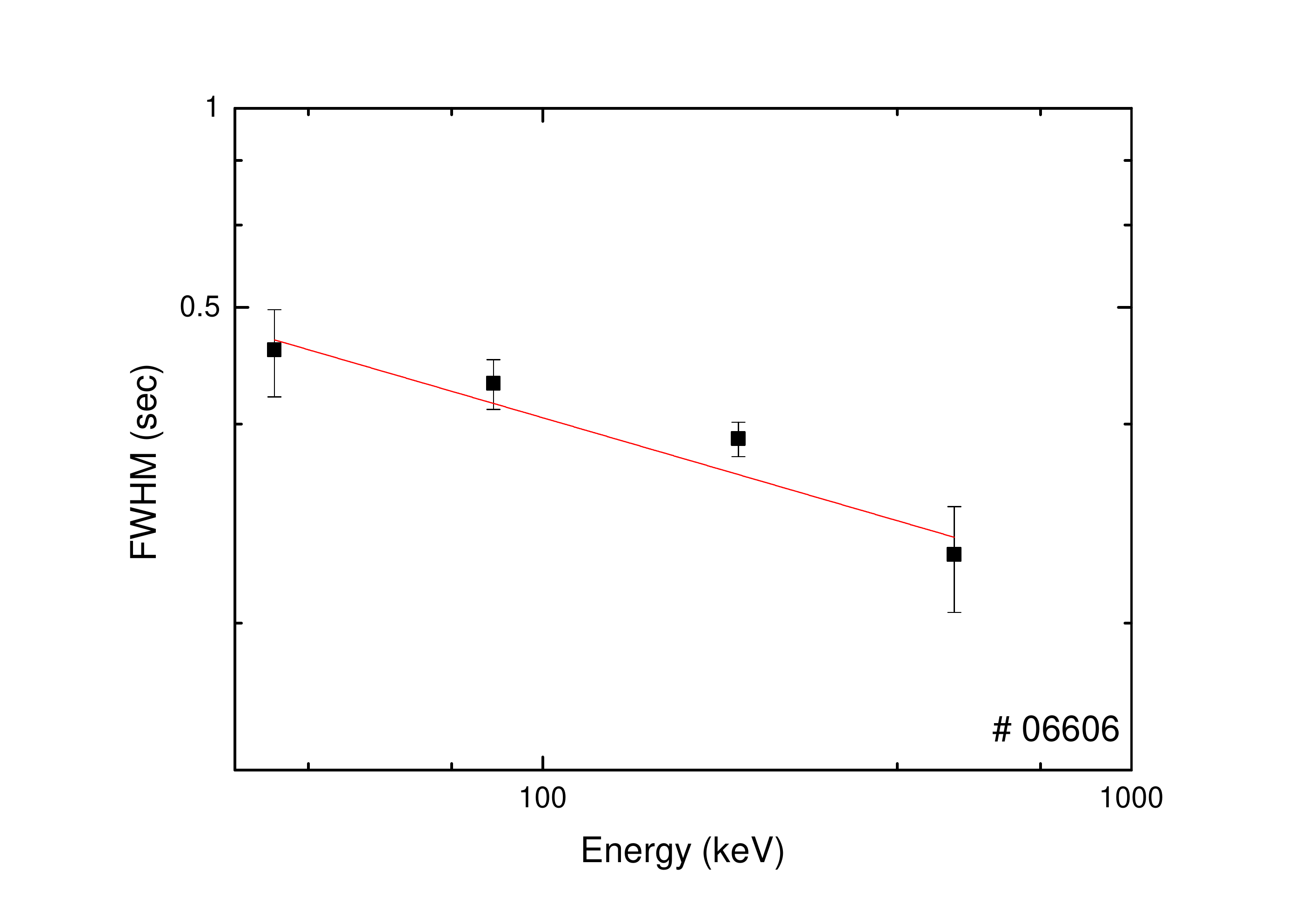}
\end{minipage}
}
\vspace{-0.1cm}
\subfigure{
\begin{minipage}{5cm}
\centering
\includegraphics[width=5cm]{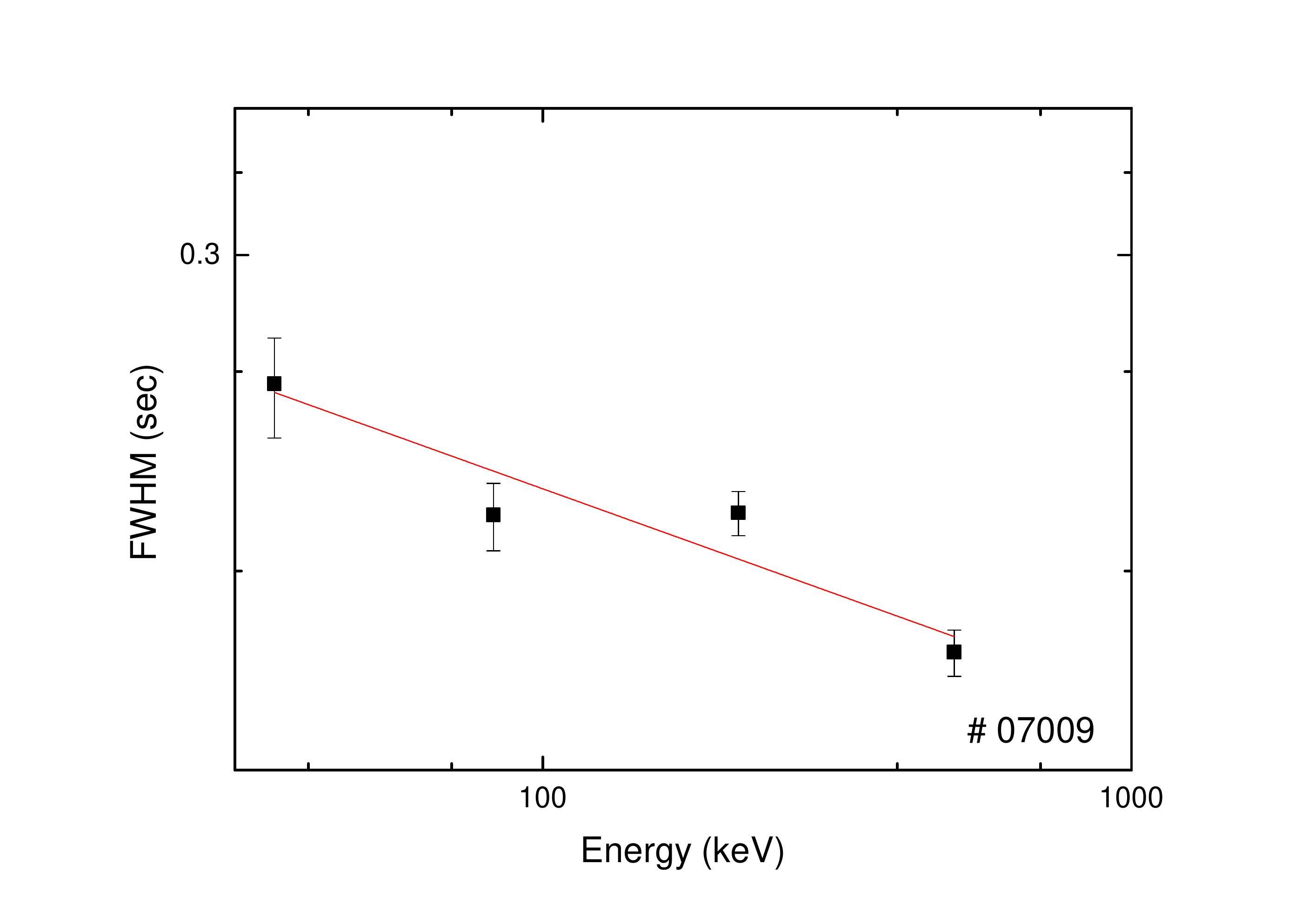}
\end{minipage}
}
\vspace{-0.1cm}
\subfigure{
\begin{minipage}{5cm}
\centering
\includegraphics[width=5cm]{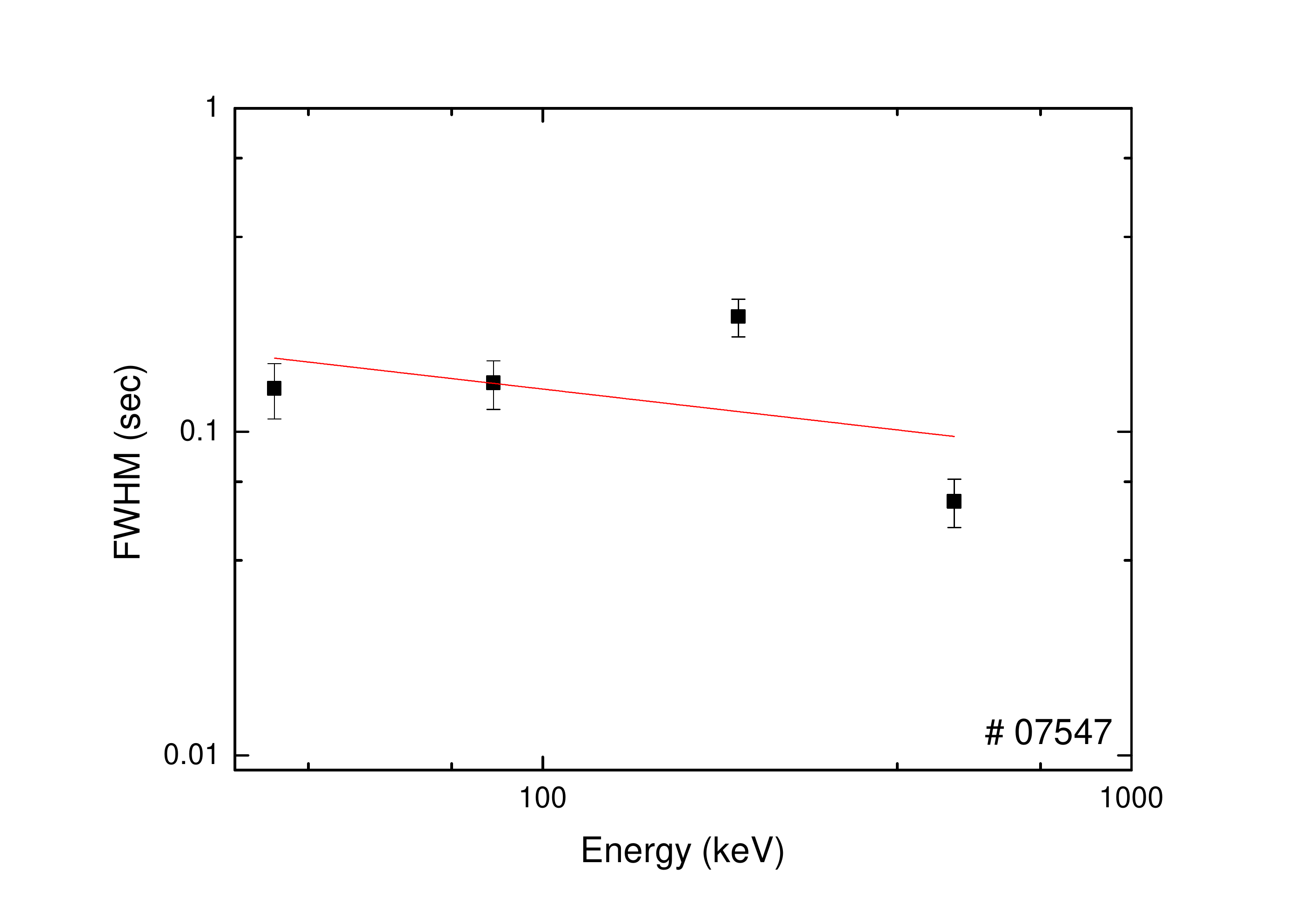}
\end{minipage}
}
\caption{Examples of the dependence of the FWHM on the photon energy with a power law anti-correlation for the SPs. Note that the inferred errors of the FWHM for
the pulses in Ch4 of $\#$ 02068 and in Ch3 of $\#$ 00432 are so large that we just assume a 10\% of the FWHM as an error estimation in our calculations, as reported in \cite{1996ApJ...459...393} and \cite{2005ApJ...627...324}.
}
\label{FWHMandesingle1-figure12}
\end{figure*}

\begin{figure*}
\centering
\gridline{
\fig{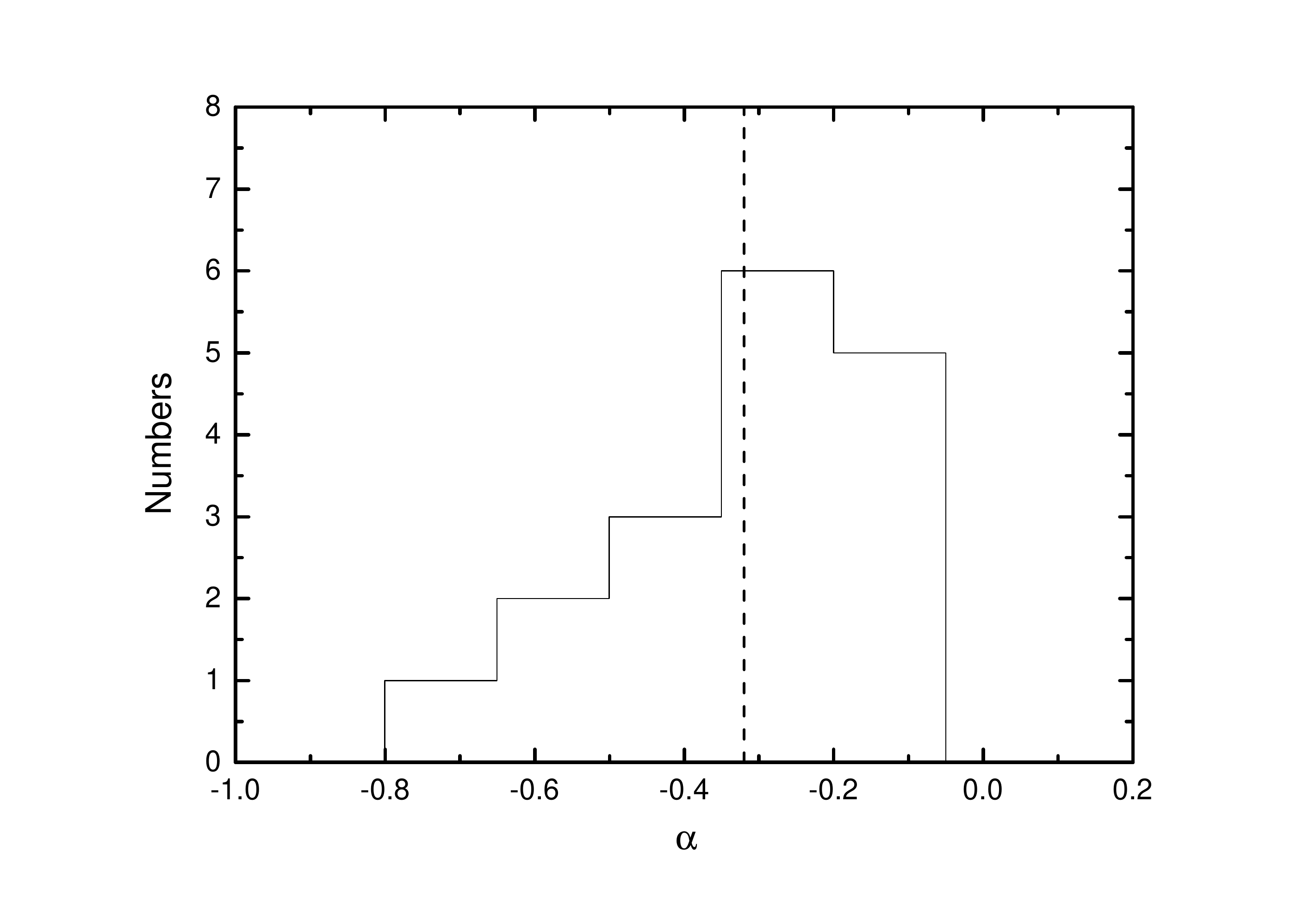}{0.6\textwidth}{}
          }
\caption{Distribution of  the $\alpha$ in the $FWHM \propto E^\alpha$ relation for 17 SPs with negative power law indexes.  The vertical dash line indicates the
mean value of $\alpha \simeq$ -0.32.
\label{FWHMandesingle1-d-figure13}}
\end{figure*}

\begin{figure*}
\centering
\gridline{
\fig{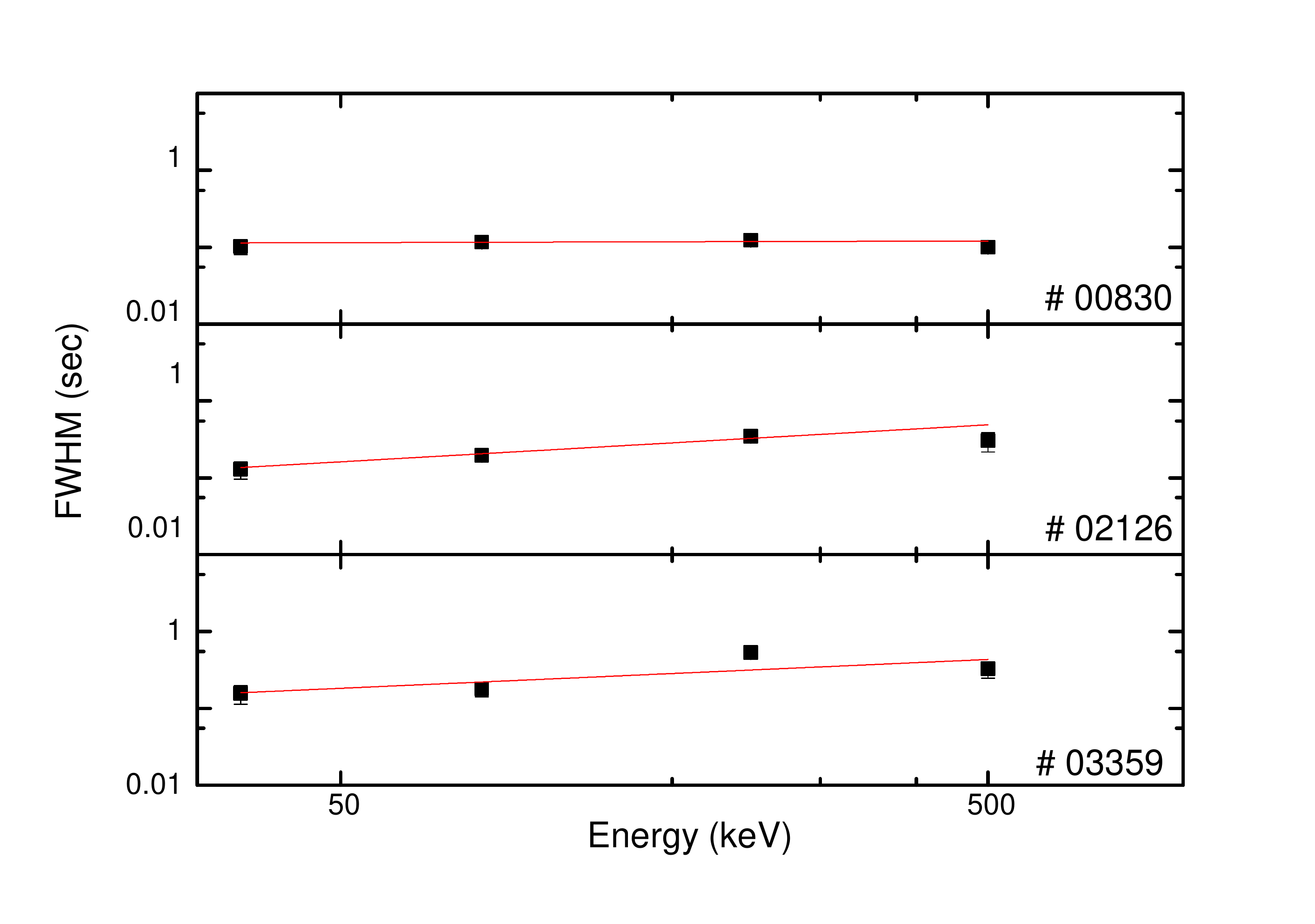}{0.6\textwidth}{}
          }
\caption{Examples of the dependence of the FWHM on the photon energy with a positive power-law correlation for the three SPs numbered with \#00830, \#02126 and
\#03359.
\label{FWHMandesingle2-figure14}}
\end{figure*}

\begin{figure*}[ht]
\centering
\gridline{
\fig{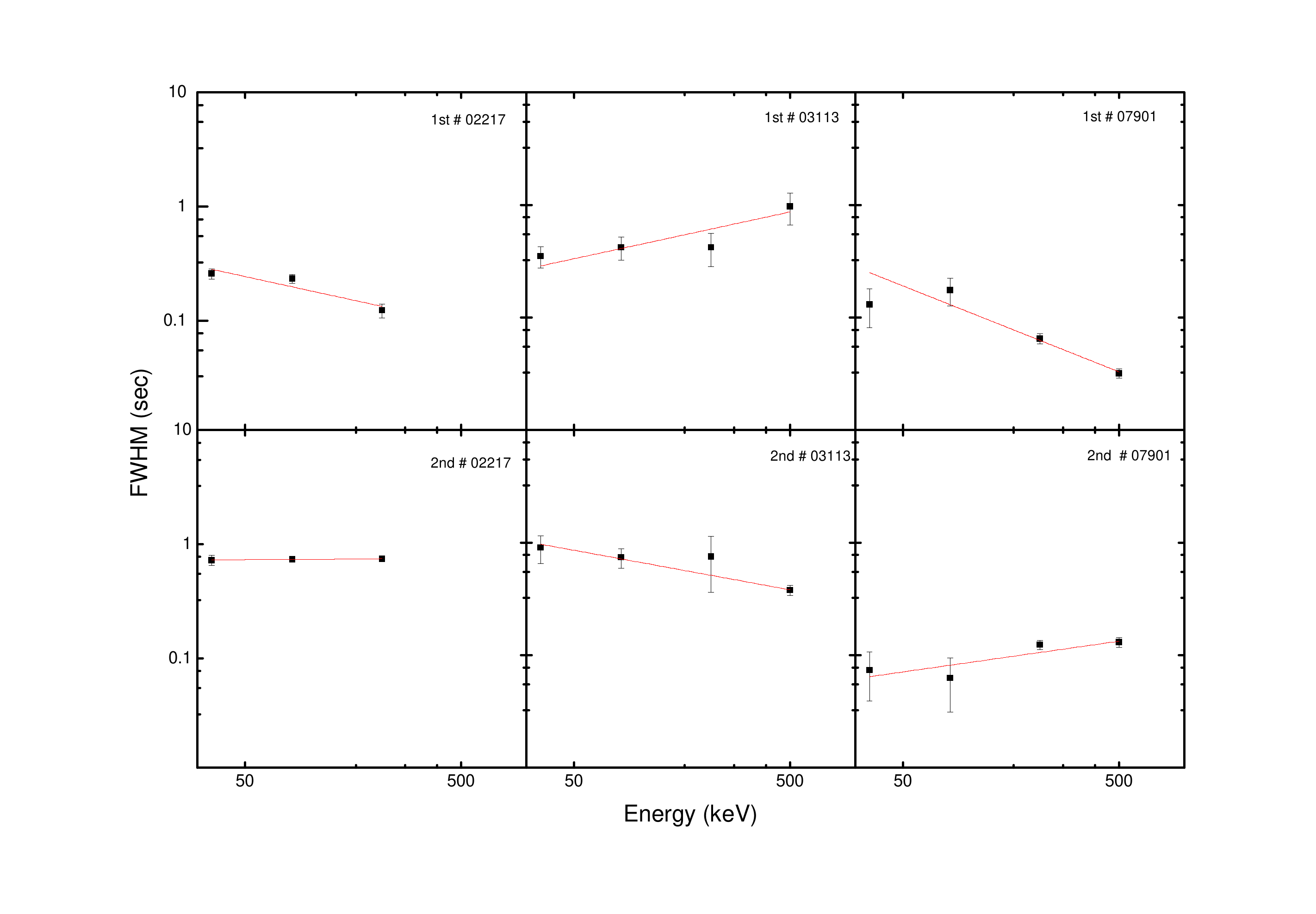}{0.45\linewidth}{(a)}
\fig{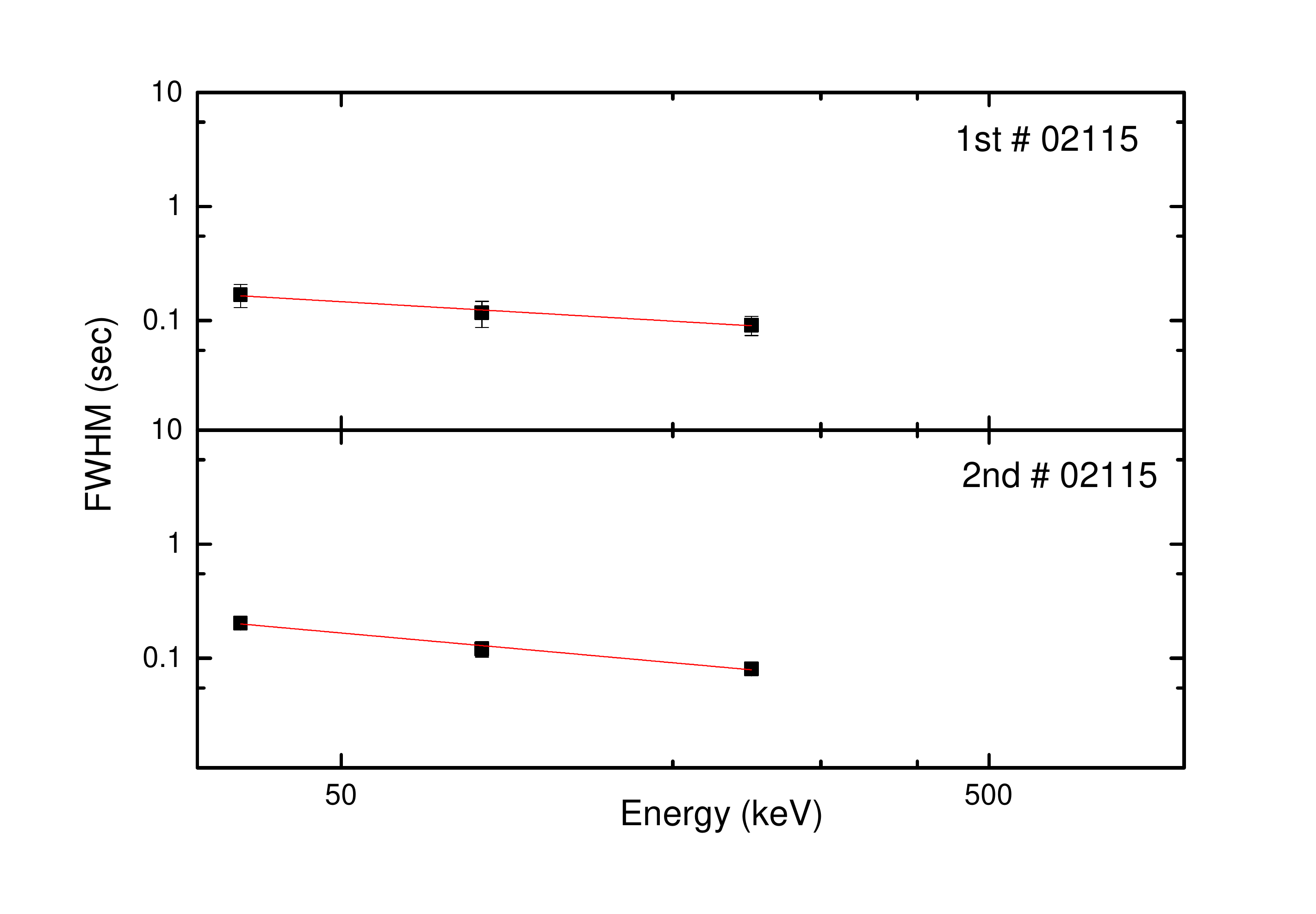}{0.45\linewidth}{(b)}
}
\caption{ (a) The dependencies of the FWHM on the photon energy for the Mt-DP sGRBs. Note that the inferred errors of the FWHM for the 1st and the 2nd pulses in
Ch2 of \# 02217, Ch1 and Ch3 of \# 07901 and the 2nd pulse in Ch4 of \# 03113 are so large that we just take a 10\% of the FWHM as an error estimation in our
calculations. (b) The dependencies of the width on the photon energy for the first (top panel) and the second (bottom panel) pulses of the Ml-DP sGRB (\# 02115).
Note that the inferred errors the FWHM in the second pulse are too large hence we instead take 10\% of the widths as the estimated errors at each point, as reported in \cite{1996ApJ...459...393} and \cite{2005ApJ...627...324}.
\label{FWHMande-figure15}}
\end{figure*}

Figure \ref{FWHMandesingle1-figure12} $-$ ~\ref{FWHMandesingle1-d-figure13} display the fitted 17 SPs with negative power-law indexes. The mean value of the
negative indexes is $\alpha\simeq-0.32\pm0.03$ that is very close to the results gotten by \cite{1995APJ...448...L101} and \cite{1996ApJ...459...393} for the
pulses of long bursts. On the other hand, we surprisingly find that three SP sGRBs have positive power-law indexes whose mean value is about
$\alpha\simeq0.29\pm0.09$  as shown in Figure \ref{FWHMandesingle2-figure14}. In addition, we select four DPs to compare with the above results of the SPs in
Figure \ref{FWHMande-figure15} (a) and (b), where it can be seen that the power law indexes in the first and the second pulses of the Mt-DPs, unlike the those of
the Ml-DPs, are inverse and inconsistent. The mean values of these negative and positive power-law indexes are $-0.49\pm0.14$ and $0.19\pm0.05$, respectively for
the 1st and the 2nd pulses in the Mt-DPs. For the Ml-DPs, both two pulses of $\#$ 02115 have a mean index value of $\alpha\simeq-0.42\pm0.03$ that is well
consistent with that of long GRBs. \cite{2006MNRAS...368...1351} studied two samples of long BATSE bursts taken from \cite{2003APJ...596...389} and
\cite{1999APJ...518...901} and found that five bursts behave a positive index power-law correlation between the FWHM and the photon energy. They pointed out that
the formation mechanism of the positive correlation was unknown. Very excitingly, we claim that the positive power law correlations not only exist in long GRBs,
but also appear in some of the SPs and the DPs.

Interestingly, \cite{Norrisetal2011} studied the heterogeneity of \textit{Swift}/BAT sGRBs and found that pulse peak intensity and pulse interval or width are strongly (weakly) anti-correlated with each other for sGRBs without (with) extended emission (EE) components or for some lGRBs \citep{Hakkila2008,2011ApJ...740...104}, which means the pulse width and the interval are proportionally related for both short and long bursts, at least parts of them. Since the pulse intensity is tightly relevant to the total $\gamma$-ray energy, the negative correlation of $FWHM\propto E^{\alpha}$ is naturally expected. Zhang et al. (2007) proposed that short and long GRBs respectively occurred at smaller and larger radii from their central regions \citep{2007AN...328...99}. If the pulse timescale is mainly contributed by the accretion process to an incipient black hole \citep{Norrisetal2011}, the inconformity of pulse emission strength (or energy), width or interval between two kinds of bursts can resulted from their different emitting radii.

\subsection{Temporal Evolution}
\begin{figure*}
\centering
\gridline{
\fig{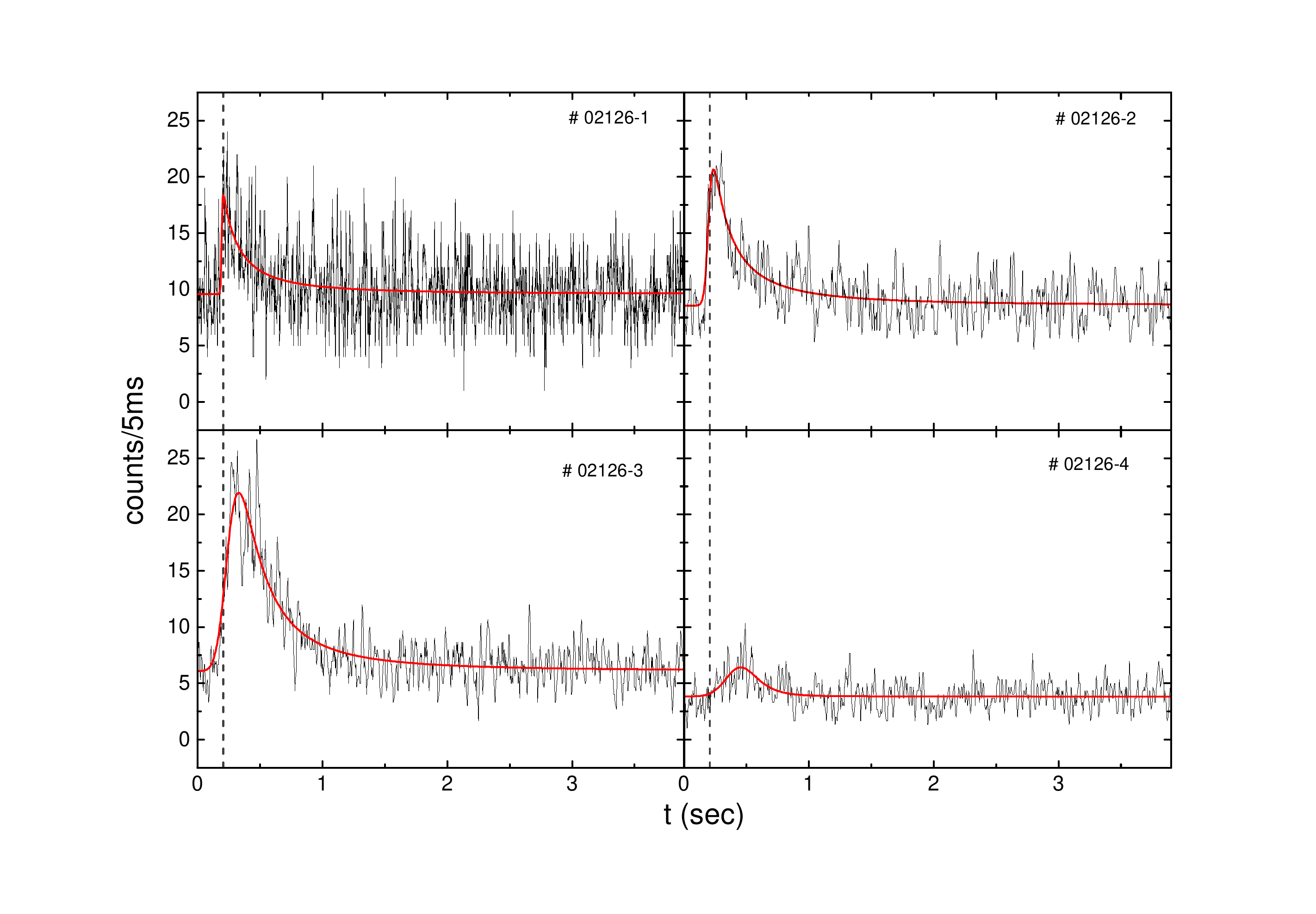}{0.7\textwidth}{}
          }
\caption{The pulse shape revolutions of the trigger \# 02126 through Ch1 (upper left panel), Ch2 (upper right panel) , Ch3 (lower left panel) and Ch4 (lower right
panel). The peak time of pulse in Ch1 is marked by the vertical black dash line. Note that the dependence of the FWHM on the photon energy has a positive
power-law index of $\alpha = 0.48\pm0.13$.
\label{02126pulse-figure17}}
\end{figure*}

\begin{figure*}
\centering
\gridline{
\fig{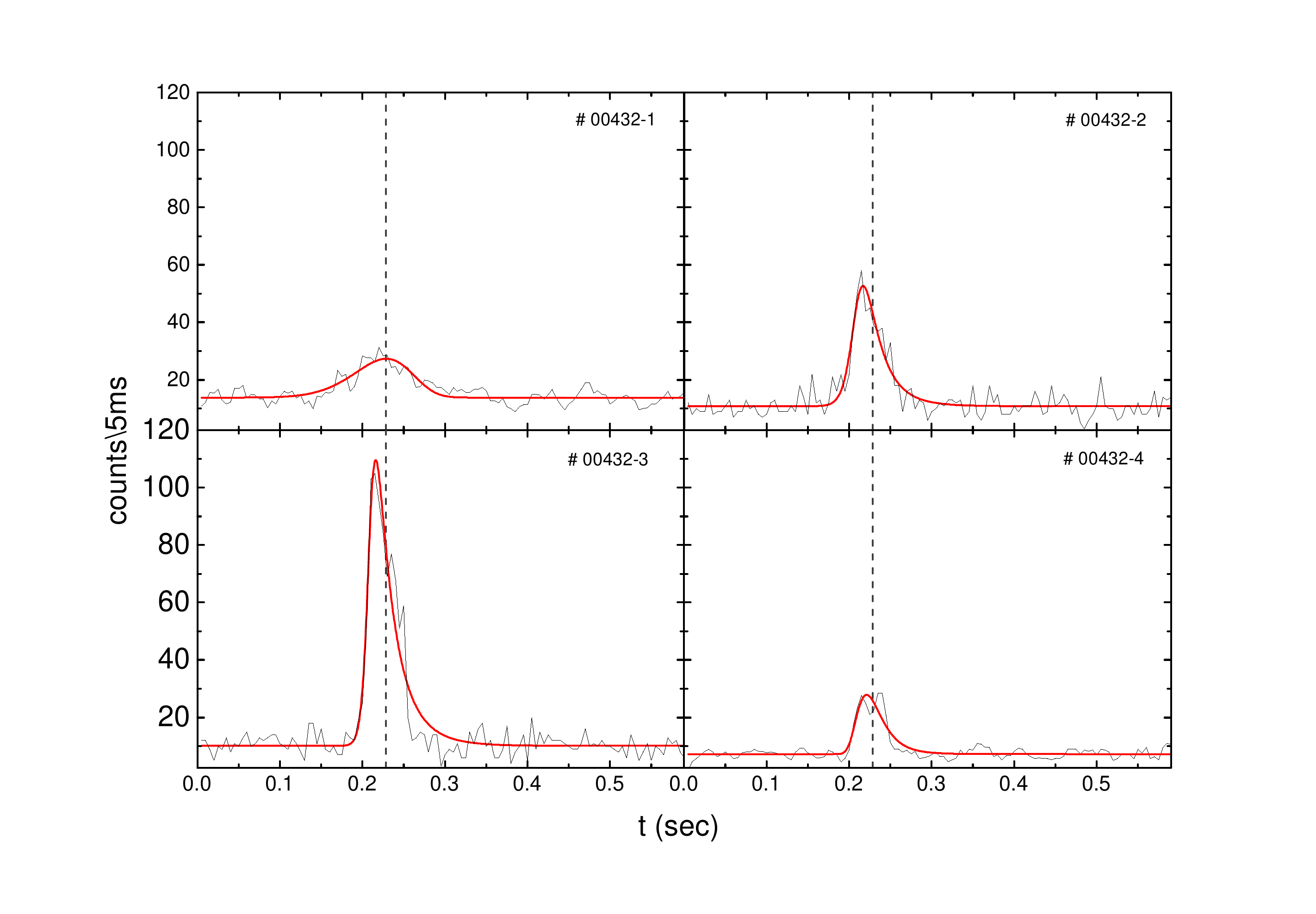}{0.7\textwidth}{}
          }
\caption{The pulse shape revolutions for the trigger \# 00432 through Ch1 (upper left panel), Ch2 (upper right panel), Ch3 (lower left panel) and Ch4 (lower right
panel). The peak time of pulse in Ch1 is marked by the vertical black dash line. Note that the dependence of the FWHM on the photon energy has a negative
power-law index of $\alpha = -0.32\pm0.15$.
\label{00432pulse-figure18}}
\end{figure*}

Owing to the joint effects of radiation processes, geometries and dynamics of outflows, the observed temporal profiles usually evolve with both the time and the
frequency in the observer frame. Figures \ref{02126pulse-figure17} and ~\ref{00432pulse-figure18} exhibit two typical examples with temporal evolutions from lower
to higher energy channels. We can see from Figure \ref{02126pulse-figure17} that $\#$ 02126, as one of three SPs with positive power-law index in the relation of
$FWHM\propto E^{\alpha}$, obviously changes from the asymmetrical shape in lower energy band to the symmetrical one in the higher energy band. The values of
t$_r$/$t_d$ are 0.12, 0.31, 0.49 and 0.88 in channels 1, 2, 3 and 4, respectively. Hereafter, we call this evolution sequence as ``MODE I''. On the contrary, $\#$
00432 as another SP sGRB has a common negative power-law index, $\alpha$, and evolves in shape in an opposite way roughly, which is defined to be ``MODE II''. The
values of t$_r$/$t_d$ of $\#$ 00432 from the lower to the higher channel are 1.24, 0.64, 0.56, and 0.66 corresponding to channels 1-4 individually. Furthermore,
we  analyze the evolution behavior of the first pulses of the Mt-DPs and find that the temporal profiles of $\#$ 02217 and $\#$ 07901 with a negative index
$\alpha$ are much similar to the SPs becoming more and more symmetrical. However,  the 1st pulse of $\#$ 03113 with a positive index $\alpha$ is different from
$\#$ 02126 and $\#$ 00432.

\section{IMPLICATIONS}
\begin{figure*}
\centering
\gridline{
\fig{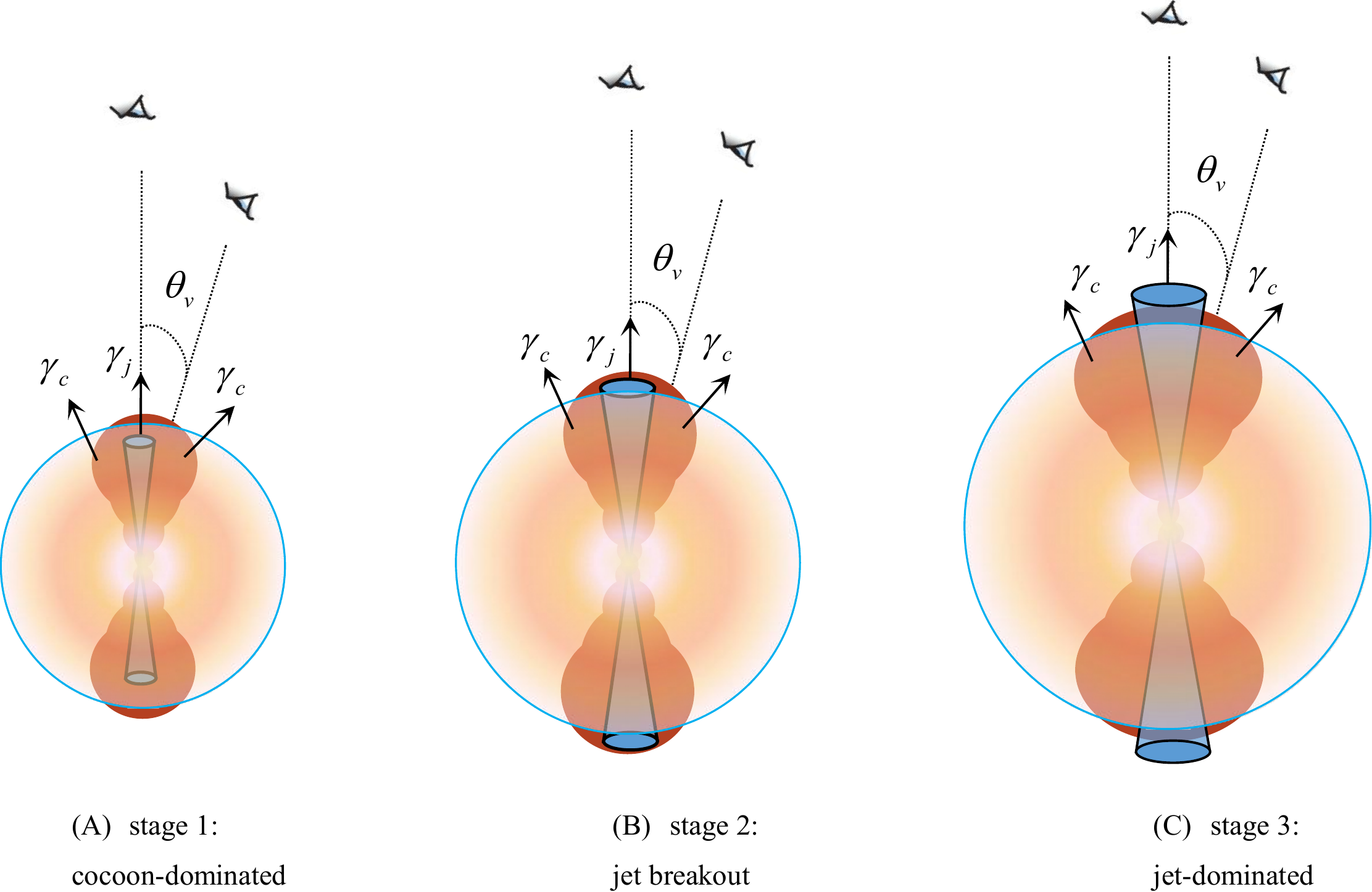}{0.7\textwidth}{}
          }
\caption{Schematic diagrams showing the geometry and evolution of a relativistically dynamical jet surrounded by a subrelativistic cocoon in some sGRBs at
different prompt $\gamma$-ray phases. It has been assumed that the following three scenarios will continuously occur, in detail, the stage 1 is dominated by the
cocoon, the stage 2 corresponds to the jet breakout, and the relativistic jet will be dominant in the stage 3.
\label{cocoon-figure19}}
\end{figure*}

In this section, we will focus on how to explain the regular evolution of two kinds of GRB pulses across different energy channels. The varieties of GRB pulses
can naturally reflect the activities of central engine, together with the geometry of the jetted outflows. As mentioned in \cite{2007AN...328...99}, the temporal
profiles of GRB pulses are usually determined by three main timescales, namely the angular spreading timescale, T$_{ang}=R_c/(2\Gamma^2c$), the dynamic timescale,
T$_{dyn}=\Delta'/(2\Gamma \upsilon_{sh}')$ and the cooling timescale, T$_{syn}=t'_{\gamma}/\Gamma$, where $R_c$ is the radius of emission region, $\Gamma$ is the
bulk lorentz factor of the outflow, $c$ is the speed of light in vacuum, $\Delta'$ is the thickness of shell and $\upsilon'_{sh}$ is the velocity of shock
relative to the pre-shocked flow, and $t'_{\gamma}$ is the radiative timescale in co-moving frame of the shell. In different phases after a burst, the above three
timescales will change with time or radius. For instance, the cooling timescale T$_{syn}$ possibly becomes longer than the other two timescales at a larger radius
so that the resulting pulse profiles could behave a quasi-FRED but more symmetric feature with smoother peaks \citep{2007AN...328...99,2000APJ...537...824}.
 In other words, the pulse shapes would be dominated by parts of these timescales in a certain emitting region. For simplicity, we assume that one GRB pulse has
 been generated from internal collision shocks within a two-component jet as illustrated in Figure \ref{cocoon-figure19}, where $\gamma_j$ and $\gamma_c$ are
 lorentz factors of the inner jet and the exterior cocoon, respectively. Besides, the $\gamma_j$ is always far larger than the $\gamma_c$ in any cases. We also
 caution that the outward cocoon around the jet may have a sub-relativistic velocity, instead of the non-relativistic one inferred for GRB 170817A \citep
 {2018Nature...554...207,2018Nature...561...355,2019Science...363...968}.

The temporal evolution ``MODE I'' can be easily explained if the Figure \ref{cocoon-figure19} is considered on-axis as follows. At early stage 1, the cocoon
envelope ahead of the jet surface will be accelerated by the inner jet at larger angle range due to the viscidity of the outflows. The earlier soft $\gamma$-rays
are mainly produced from the sub-relativistic cocoon that is heavily suffered from the curvature effect. This will result in the FRED-like pulses as we have seen
in lower energy channels, for instance the upper left and right panels of Figure \ref{02126pulse-figure17}. Since the inner jet moves outwards far faster than the
outer cocoon, the jet will pass through the cocoon and start to contribute $\gamma$-rays dominantly to the GRB pulse in higher energy channels from the stage 2 to
3. At this moment, the temporal profiles dominated by higher energy photons are primarily resulted from the dynamic and synchrotron timescales in that the newborn
jet is much narrower than the forgoing cocoon, which leads to the observed pulses become more and more symmetrical. Of cause, if the jet-cocoon system is seen at
a off-axis angle of $\theta_{v}$, the spreading angular timescale will play more important role on shaping the pulses.

 Regarding the ``MODE II'', we intend to interpret it by taking into account an isolate jet without a cocoon accompanied. In fact, for most of sGRBs, the
surrounding cocoons are not always necessary on modeling the potential progenitors. Here, we assume that the cocoon in Figure \ref{cocoon-figure19} does not exist
from the beginning and prompt $\gamma$-rays are only emitted from the relativistic jet. In early phase after the jet was launched, its initial opening angle is
too narrow to contribute more $\gamma$-rays influenced by the curvature effect significantly. Thus the early pulses in lower energy channels would be mainly
determined by the dynamic and synchrotron timescales and behave more symmetrical. With the processes of the acceleration of the jet and its interaction with
circum-burst medium, the effect of sideways expanding \citep{1999ApJ...525...737,1999ApJL...524...L43} on the spreading angular timescales should be more
significant than before. Naturally, the observed pulses in this situation will have the FRED-like shapes as displayed in Figure \ref{00432pulse-figure18}. In
addition, the trend of the FRED will be strengthened if the jet is viewed from a larger off-axis angle.

\section{CONCLUSIONS AND DISCUSSIONS}
In this paper, we have systematically studied the statistical properties of short BATSE GRBs and achieve the following conclusions.

\begin{itemize}
	\item There is a consistent power-law correlation between the rising time t$_r$ and the decay time t$_d$ for the SPs and the first/second pulse of the DPs,
except the second pulse of the Ml-DPs, which may suggest that two types of DPs might be originated from different physical processes. Also, we verify that the
distributions of the asymmetry in the SPs and the first pulses of the DPs are drawn from the same parent distribution. The mean value of the asymmetry $t_r/t_d$
of the SPs is consistent with some previous results gotten by \cite{2007APSS...310...19}, showing again the pulses in sGRBs are more symmetrical than long GRB
pulses.

	\item On the whole, there are no obvious relationships of the asymmetry with pulse width, the peak time t$_m$ and the amplitude f$_m$, respectively, found
for the SPs and the DPs of the sGRBs, except the second pulses in the Ml-DPs that behave weak dependence of the $t_r/t_d$ on the FWHM and the t$_m$. This may
hint that the Ml-DP indeed has a unique origin. The result of the asymmetry uncorrelated with the width is coincident with the previous finding for long GRBs by
\cite{2003APJ...596...389}.

	\item We have studied the f$_m$ distributions contrastively and concluded that the averaged magnitude of the SPs is clearly higher than those of the
first/second pulses in the DPs. It is surprisingly found that the released energy amount in the 1st pulses of the Mt-DPs is larger than that in the 2nd pulses,
while it is opposite for the Ml-DPs, which may demonstrate again that these two kinds of DPs are largely different in physical processes.

    \item To compare the power-law relation, $FWHM\propto E^{\alpha}$, of the photon energy with the pulse width of the sGRBs with that discovered for long GRBs
        previously, we chose both the SPs and the DPs as our research targets. Interestingly, not only the traditionally negative but also the peculiarly
        positive energy correlations are found to coexist in either the SPs or the DPs. We noticed that the negatively mean index of $< \alpha> \sim -0.3 $ for
        the SPs is very close to some previous results of long GRBs \citep [e.g.][]{1995APJ...448...L101, 1996ApJ...459...393}. For the DPs, it is hard to
        obtain the reliable power law index $\alpha$ statistically due to the currently limited samples.

    \item Finally, we studied the evolution of different kinds of pulses in sGRBs across diverse energy channels and found two regular evolving modes, namely
        the ``MODE I'' and the ``MODE II'', for the first time. Importantly, we speculated that the two evolving modes of pulses could universally appear in
        most of GRBs. More importantly, we proposed that the regular types of ``MODE I'' and the ``MODE II'' can be used as a probe to explore the structures,
        evolutions and orientations, etc, of the relativistic outflows, especially in cases of the off-axis and lateral expansion. However, the current samples
        with the two precise modes are relatively small in the BATSE catalog. It is vastly encouraged to search for more samples with the standard modes from
        other GRB catalogs, such as Fermi/GBM, HXMT/HE, \textit{Swift}/BAT.
\end{itemize}

Some similarities of long and short GRBs in respect of the observational properties of prompt gamma-rays had been shown in \citep{2015JHEA...7...81}. Regarding
the softer X-rays, interestingly, it had been pointed out by \cite{2011MNRAS...417...2144} that the early X-ray flares of short GRBs behaved as those of long
GRBs. Recently, Hakkila et al. \citep{2014ApJ...783...88,2015ApJ...815...134,2018ApJ...855...101} examined the residuals in their fits to long, intermediate and short
GRB pulses and found three separate wave-like peak structures in the residuals for each pulse. Then they suggested that the complex GRB profiles may be composed of fewer
pulses than the apparent number of peaks. This might help us to understand the diversity of multiple pulses in some bursts, other than the aspect of physical
mechanisms.

The internal energy dissipation processes of GRBs include the precursors emissions caused by shock breakout or photosphere emissions
\citep{2000ApJL...543...L129,2006Nature...442...1008}, prompt $\gamma$-ray emissions (the main emission), EE as well as late X-ray flares
\citep{2014ApJ...789...145}. The precursors and the EE had been investigated by many previous authors
\citep[e.g.,][]{2000ApJL...543...L129,2006Nature...442...1008,2018Nature...2...69,2016ApJ...829...7,2018ApJ...862...155,2019Zhong}. Similarly, the precursors and
the EE components might be physically related with the main radiations of $\gamma$-rays and are expected to disclose more progenitor information for the sGRBs (Li
et al. 2020, in preparation). For brighter sGRBs, the EE component is sometimes detectable above the background \citep{Norris2006}. One SP (\#07427) and one DP (\#00575) in our sample are two EE bursts from \cite{2018ApJ...855...101} and \cite{2013MNRAS...428...1623B}. However, we cannot fit the EE episodes with the pulse model because of their low signal-to-noise as in \cite{2018ApJ...855...101}. In practice, the EE components are more readily perceived in \textit{Swift}/BAT mask-tagged data for sGRBs, we have examined in details their temporal and spectral properties and possible connections with main peak episodes (\cite{li2020}, in preparation).

Very recently, an interesting work on multiple pulses of Fermi/GBM GRBs had been done by \cite{2019ApJs...242...16} to search for an
evidence of the transition from fireball to Poynting-flux-dominated outflow in the GRB160602B-like sample. As a result, he found 9 out of 41 GBM bursts to be
coincident with the case. Note that multiple pulses in each of the 41 bursts can be clearly separated. Actually, the prompt $\gamma$-rays emitted from smaller to
larger radii should be naturally composed of different radiative components. The reason is that the early and late dominant radiation mechanisms are generally the
thermal emissions from the photospheres and the non-thermal (synchrotron) emissions from the Poynting fluxes. According to \cite{2019ApJs...242...16}, the
differences between the Ml-DPs and other sGRBs with the SPs and the Mt-DPs can be easily explained if the Ml-DPs are assumed to occur the transition process from
fireball to Poynting-flux-dominated outflow. Based on the above discussions, we believe that the observed profiles of GRB pulses are affected by not only the
geometry, the dynamics and the radiative cooling of outflows, but also the detailed radiation mechanisms. Certainly, the distinct radiation mechanisms will lead
to different radiative cooling processes. Therefore, it is very urgent to clarify the fraction of the SPs belonging to the MODEs I and II, and distinguish how the
diverse radiation mechanisms affect on the shape of all kinds of pules in details in the future.

Acknowledgements

We are very grateful to the anonymous referee for his/her comments, which helped us to improve the manuscript greatly. This work made use of the data supplied by the High Energy Astrophysics Science Archive Research Center (HEASARC) for the online CGRO/BATSE catalog. It was partly
supported by the Natural Science Foundations (ZR2018MA030, XKJJC201901).

\clearpage
\startlongtable
\begin{deluxetable*}{cccccc}
\tabletypesize{\small}
\tablecaption{The best-fit parameters of the power-law correlation between the t$_r$ and the t$_d$.\label{table1}}
\tablehead{
\emph{GRB}&$\emph{Class}$&$\emph{$\chi^2_\nu$}$& $\emph{Pearson's r}$&\emph{R}& $\emph{$\beta$}$}
\startdata
SPs&Ch2&3.64 &0.80 &0.64&$0.89 \pm 0.09$\\
SPs&Ch3&1.54 &0.92 &0.85&$1.04 \pm 0.06$\\
Mt-DPs$^\dag$&Ch2/3 &1.20 &0.86 &0.73&$1.10 \pm 0.13$\\
Mt-DPs$^\ddag$&Ch2/3 &1.39 &0.84 &0.69&$0.83 \pm 0.11$\\
Ml-DPs$^\dag$&Ch2/3&0.10 &0.57 &0.23&$0.49 \pm 0.27$\\
Ml-DPs$^\ddag$&Ch2/3&0.14 &0.02 &-0.14&$0.01 \pm 0.19$\\
\enddata
\end{deluxetable*}

\startlongtable
\begin{deluxetable*}{ccccc}
\tabletypesize{\small}
\tablecaption{Asymmetric properties of different sGRBs.\label{table2}}
\tablehead{
\emph{GRB}&\emph{Number}&\emph{Maximum}&\emph{mean(Median)} & \emph{Minimum}}
\startdata
SPs&Ch2+3&1.28&0.73(0.76)&0.04\\
Mt-DPs&1st & 1.23&0.82(0.82)&0.50\\
&2nd & 1.01&0.51(0.49)&0.12\\
Ml-DPs&1st &1.43&0.80(0.84)&0.20\\
&2nd &1.48& 1.07(1.21)&0.16\\
Total  & -&1.48&0.79(0.81)&0.04\\
\hline
TPs   &1st&-&1.06&-\\
($\#$ 06715)   &2nd&-&0.57&-\\
              &3th&-&1.39&-\\
 TPs  &1st&-&0.56&-\\
 ($\#$ 07102)  &2nd&-&0.54&-\\
              &3th&-&0.93&-\\
\enddata
\end{deluxetable*}

\startlongtable
\begin{deluxetable*}{cccccc}
\tabletypesize{\small}
\tablecaption{The best-fit parameters of the correlation of the t$_m$ with the FWHM. \label{table3}}
\tablehead{
\emph{GRB}& $\emph{Class}$&$\emph{$\chi^2_\nu$}$&$\emph{Pearson's r}$ & \emph{R}& $\emph{$\mu$}$}
\startdata
SPs&Ch2&0.78&0.73&0.52&$0.56 \pm 0.07$\\
SPs&Ch3&0.75&0.72&0.51&$0.61 \pm 0.08$\\
DPs&Mt-DPs 1st&0.19&0.69 &0.45&$0.62 \pm 0.13$\\
&Mt-DPs 2nd&0.55&0.91&0.81&$1.33 \pm 0.12$\\
DPs&Ml-DPs 1st&0.01&0.35 &0.00&$0.54 \pm 0.54$\\
&Ml-DPs 2nd&0.13&0.44 &0.08&$0.99 \pm 0.77$\\
\hline
\enddata
\end{deluxetable*}

\startlongtable
\begin{deluxetable*}{cccccc}
\tabletypesize{\small}
\tablecaption{The best-fit parameters of the correlation of the f$_m$ with the FWHM. \label{table4}}
\tablehead{
\emph{GRB}& $\emph{Class}$&$\emph{$\chi^2_\nu$}$&$\emph{Pearson's r}$ & \emph{R}& $\emph{$\nu$}$}
\startdata
SPs&Ch2&3.63&-0.64&0.40&$-0.57 \pm 0.09$\\
SPs&Ch3&3.52&-0.73&0.53&$-0.69 \pm 0.09$\\
DPs&Mt-DPs 1st&1.18&-0.54&0.27&$-0.39 \pm 0.12$\\
&Mt-DPs 2nd &1.80&-0.38&0.11&$-0.27 \pm 0.13$\\
DPs&Ml-DPs 1st&0.41&-0.09&-0.13&$-0.16 \pm 0.65$\\
&Ml-DPs 2nd&0.12&-0.79&0.58&$-0.80 \pm 0.23$\\
\hline
\enddata
\end{deluxetable*}

\startlongtable
\begin{deluxetable*}{clccccc}
\tabletypesize{\small}
\tablecaption{ The best-fit parameters of the correlation between the FWHM and the average energy of photons in each channel.\label{table5}}
\tablehead{
\emph{GRB}&\emph{Trigger number}&\emph{Class}&\emph{$\alpha$}& \emph{$\chi^2_\nu$} &\emph{Pearson's r}& \emph{R} }
\startdata
SPs&00432&negative&${-0.32\pm0.15}$& 12.50 &-0.83&0.53\\
SPs&00474&negative&${-0.10\pm0.04}$& 3.09 &-0.89&0.69\\
SPs&00480&negative&${-0.41\pm0.03}$& 0.29 &-1.00&0.99\\
SPs&00568&negative&${-0.40\pm0.10}$& 3.35 &-0.94&0.84\\
SPs&01076&negative&${-0.18\pm0.12}$& 2.00 &-0.74&0.32\\
SPs&01097&negative&${-0.54\pm0.18}$& 1.57 &-0.90&0.72\\
SPs&01102&negative&${-0.22\pm0.06}$& 2.39 &-0.94&0.81\\
SPs&01359&negative&${-0.25\pm0.05}$& 0.15 &-0.96&0.89\\
SPs&02068&negative&${-0.14\pm0.01}$& 0.5 &-0.83&0.53\\
SPs&02614&negative&${-0.36\pm0.07}$& 7.82 &-0.96&0.89\\
SPs&02896&negative&${-0.18\pm0.09}$& 1.91 &-0.89&0.58\\
SPs&03737&negative&${-0.77\pm0.06}$& 0.51 &-0.99&0.98\\
SPs&03940&negative&${-0.11\pm0.03}$& 0.16 &-0.94&0.82\\
SPs&04955&negative&${-0.64\pm0.11}$& 1.30 &-0.97&0.91\\
SPs&06606&negative&${-0.26\pm0.04}$& 2.34 &-0.97&0.92\\
SPs&07009&negative&${-0.32\pm0.08}$& 3.14 &-0.95&0.84\\
SPs&07547&negative&${-0.21\pm0.30}$& 13.53 &-0.44&-0.20\\
\hline
SPs&00830&positive&${0.02\pm0.07}$& 0.91 &0.21&-0.43\\
SPs&02126&positive&${0.48\pm0.13}$& 2.66 &0.93&0.80\\
SPs&03359&positive&${0.38\pm0.22}$& 19.36 &0.77&0.38\\
\hline
Mt-DPs&02217 1st&negative&${-0.41\pm0.16}$& 4.29 &-0.93&0.74\\
Mt-DPs&02217 2nd&positive&${0.02\pm0.01}$& 0.02 &0.95&0.79\\
Mt-DPs&03113 1st&positive&${0.42\pm0.15}$& 1.35 &0.89&0.68\\
Mt-DPs	&03113 2nd& negative&${-0.35\pm0.04}$& 0.24 &-0.99&0.96\\
Mt-DPs	&07901 1st&negative&${-0.72\pm0.40}$& 3.60 &-0.79&0.43\\
Mt-DPs	&07901 2nd &positive&${0.12\pm0.01}$& 1.31 &0.99&0.96\\
\hline
Ml-DPs&02115 1st&negative&${-0.33\pm0.04}$& 1.50 &-0.99&0.97\\
Ml-DPs	& 02115 2nd&negative&${-0.51\pm0.03}$& 0.11 &-1.00&0.99\\
\enddata
\end{deluxetable*}


\begin{thebibliography}{}
\expandafter\ifx\csname natexlab\endcsname\relax\def\natexlab#1{#1}\fi
\providecommand{\url}[1]{\href{#1}{#1}}
\providecommand{\dodoi}[1]{doi:~\href{http://doi.org/#1}{\nolinkurl{#1}}}
\providecommand{\doeprint}[1]{\href{http://ascl.net/#1}{\nolinkurl{http://ascl.net/#1}}}
\providecommand{\doarXiv}[1]{\href{https://arxiv.org/abs/#1}{\nolinkurl{https://arxiv.org/abs/#1}}}

\end{thebibliography}


\clearpage
\begin{thebibliography}{}
\bibliographystyle{phjcp}

\bibitem[Beniamini \& Nakar(2019)]{2019MNRAS...482...5430} Beniamini, P., \& Nakar, E. \ 2019, \ MNRAS, 482, 5430
\bibitem[Bostanc{\i} et al.(2013)]{2013MNRAS...428...1623B} Bostanc{\i}, Z. F., Kaneko, Y., \& G\"{o}\u{g}\"{u}\c{s}, E.\ 2013, \mnras, 428, 1623
\bibitem[Campana et al.(2006)]{2006Nature...442...1008} Campana, S., Mangano, V., Blustin, A. J., et al. \ 2006, \ Nature, 442, 1008
\bibitem[Fenimore et al.(1995)]{1995APJ...448...L101} Fenimore, E. E., in't Zand, J. J. M., Norris,J. P., et al. \ 1995, \ ApJL, 448, L101
\bibitem[Ghirlanda et al.(2019)]{2019Science...363...968} Ghirlanda, G., Salafia, O. S., Paragi, Z., et al. \ 2019, \ Science, 363, 968
\bibitem[Ghirlanda et al.(2015)]{2015JHEA...7...81} Ghirlanda, G., Bernardini, M. G., Calderone, G., et al. \ 2015, \ Journal of High Energy Astrophysics, 7, 81
\bibitem[Goldstein et al.(2017)]{2017APJ...848...L14} Goldstein, A., Veres, P., Burns, E., et al. \ 2017, \ ApJL, 848, L14
\bibitem[Golkhou et al.(2015)]{2015APJ...811...93} Golkhou, V. Z., Butler, N. R., \& Littlejohns, O. M. \ 2015, \apj , 811, 93
\bibitem[Hakkila et al.(2008)]{Hakkila2008}Hakkila, J., Giblin, T.W., Norris, J. P., Fragile, P. C., \& Bonnell, J. T. 2008, \ ApJL,
677, L81
\bibitem[Hakkila \& Preece(2011)]{2011ApJ...740...104} Hakkila, J., \& Preece, R. D. \ 2011, \apj , 740, 104
\bibitem[Hakkila \& Preece(2014)]{2014ApJ...783...88} Hakkila, J., \& Preece, R. D. \ 2014, \apj , 783, 88
\bibitem[Hakkila et al.(2015)]{2015ApJ...815...134} Hakkila, J., Lien, A., Sakamoto, T., et al. \ 2015, \apj, 815, 134
\bibitem[Hakkila et al.(2018a)]{2018ApJ...855...101} Hakkila, J., Horv\'{a}th, I., Hofesmann, E., et al. \ 2018a, \apj , 855, 101
\bibitem[Hakkila et al.(2018b)]{2018ApJ...863...77} Hakkila, J., Lesage, S., McAfee, S., et al. \ 2018b, \apj , 863, 77
\bibitem[Hu et al.(2014)]{2014ApJ...789...145} Hu, Y. D., Liang, E. W., Xi, S. D., et al. \ 2014, \apj , 789, 145
\bibitem[Klebesadel et al.(1973)]{1973ApJ...182...L85} Klebesadel, R. W., Strong, I. B., \& Olson, R. A. \ 1973, \ ApJL, 182, L85
\bibitem[Kocevski et al.(2003)]{2003APJ...596...389} Kocevski, D., Ryde, F., \& Liang, E. \ 2003, \apj, 596, 389
\bibitem[Kouveliotou(1994)]{1994ApJS...92...637} Kouveliotou, C. \ 1994, \apjs, 92, 637
\bibitem[Lan et al.(2018)]{2018ApJ...862...155} Lan, L., L\"{u}, H.-J., Zhong, S.-Q., et al. \ 2018, \apj, 862, 155
\bibitem[Li(2019)]{2019ApJs...242...16} Li, L. \ 2019, \ ApJS , 242, 16
\bibitem[Li et al.(2020)]{li2020}Li, X. J., Zhang, Z. B., Zhang, K., et al. 2020, \apj, in preparation
\bibitem[Lyutikov \& Usov(2000)]{2000ApJL...543...L129}  Lyutikov, M., \& Usov, V. V. \ 2000, \ ApJL, 543, L129
\bibitem[Lin et al.(2017)]{2017APJ...840...95} Lin, D.-B., Mu, H.-J., Lu, R.-J., et al. \ 2017, \ ApJ, 840, 95
\bibitem[Lien et al.(2016)]{2016ApJ...829...7} Lien, A., Sakamoto, T., Barthelmy, S. D., et al. \ 2016, \ ApJ, 829, 7
\bibitem[Liang et al.(2007)]{2007APJ...670...565} Liang, E.-W., Zhang, B.-B., \& Zhang, B. \ 2007, \apj, 670, 565
\bibitem[Mooley et al.(2018a)]{2018Nature...554...207} Mooley, K. P., Nakar, E., Hotokezaka, K., et al. \ 2018, \ Nature, 554, 207
\bibitem[Mooley et al.(2018b)]{2018Nature...561...355} Mooley, K. P., Deller, A. T., Gottlieb, O., et al. \ 2018, \ Nature, 561, 355
\bibitem[Margutti et al.(2011)]{2011MNRAS...417...2144}  Margutti, R., Chincarini, G., Granot, J., et al. \ 2011, \ MNRAS, 417, 2144
\bibitem[McBreen et al.(2001)]{2001AA...380...L31} McBreen, S., Quilligan, F., McBreen, B., et al. \ 2001, \ A\&A, 380, L31
\bibitem[Norris et al.(1996)]{1996ApJ...459...393} Norris, J. P., Nemiroff, R. J., Bonnell, J. T., et al. \ 1996, \apj, 459, 393
\bibitem[Norris et al.(1999)]{1999APJ...518...901} Norris, J. P., Bonnell, J. T., \& Watanabe, K. \ 1999, \apj, 518, 901
\bibitem[Norris et al.(2005)]{2005ApJ...627...324} Norris, J. P., Bonnell, J. T., Kazanas, D., et al. \ 2005, \apj, 627, 324
\bibitem[Norris \& Bonnell(2006)]{Norris2006} Norris, J. P., \& Bonnell, J. T. \ 2006, \apj, 643, 266
\bibitem[Norris et al.(2011)]{Norrisetal2011} Norris, J. P., Gehrels, N., \& Scargle, J. D. \ 2011, \apj, 735, 23
\bibitem[Peng et al.(2006)]{2006MNRAS...368...1351} Peng, Z.-Y., Qin, Y.-P., Zhang, B.-B., et al. \ 2006, \ MNRAS , 368, 1351
\bibitem[Qin et al.(2004)]{2004APJ...617...439} Qin, Y.-P., Zhang, Z.-B., Zhang, F.-W., et al. \ 2004, \apj , 617, 439
\bibitem[Qin(2002)]{2002AA...396...705} Qin, Y.-P. \ 2002, \ A\&A, 396, 705
\bibitem[Quilligan et al.(1999)]{1999AAS...138...419} Quilligan, F., Hurley, K. J., McBreen, B., et al. \ 1999, \ A\&AS, 138, 419
\bibitem[Quilligan et al.(2002)]{2002AA...385...377} Quilligan, F., McBreen, B., Hanlon, L., et al. \ 2002, \ A\&A, 385, 377
\bibitem[Rhoads(1999)]{1999ApJ...525...737} Rhoads, J. E. \ 1999, \ ApJ, 525, 737
\bibitem[Ryde \& Svensson(2000)]{2000APJL...529...L13} Ryde, F., \& Svensson, R. \ 2000, \ ApJL, 529, L13
\bibitem[Ryde \& Svensson(2002)]{2002APJ...566...210} Ryde, F., \& Svensson, R. \ 2002, \ APJ, 566, 210
\bibitem[Sari(1999)]{1999ApJL...524...L43} Sari, R. \ 1999, \ ApJL, 524, L43
\bibitem[Spada et al.(2000)]{2000APJ...537...824} Spada, M., Panaitescu, A., \& M\'{e}sz\'{a}ros, P. \ 2000, \apj, 537, 824
\bibitem[Shao et al.(2017)]{2017APJ...844...126} Shao, L., Zhang, B.-B., Wang, F.-R., et al. \ 2017, \apj , 844, 126
\bibitem[von Kienlin et al.(2019)]{2019ApJ...876...89} von Kienlin, A., Veres, P., Roberts, O. J., et al. \ 2019, \apj, 876, 89
\bibitem[Willingale et al.(2007)]{2007APJ...662...1093} Willingale, R., O'Brien, P. T., Osborne, J. P., et al. \ 2007, \apj, 662, 1093
\bibitem[Wu \& MacFadyen(2018)]{2018APJ...869...55} Wu, Y., \& MacFadyen, A. \ 2018, \ ApJ, 869, 55
\bibitem[Zhang(2008)]{2008APJ...685...1052} Zhang, F.-W. \ 2008, \apj , 685, 1052
\bibitem[Zhang et al.(2006)]{2006MNRAS...373...729} Zhang, Z. B., Xie, G. Z., Deng, J. G., et al. \ 2006, \ MNRAS, 373, 729
\bibitem[Zhang \& Qin(2005)]{zhang2005} Zhang, Z. B., \& Qin, Y. P. \ 2005, \ MNRAS, 363, 1290
\bibitem[Zhong et al.(2019)]{2019Zhong} Zhong, S.-Q., Dai, Z.-G., Cheng, J.-G., et al. \ 2019, \apj, 884, 25
\bibitem[Zhang et al.(2007)]{2007AN...328...99} Zhang, Z. B., Xie, G. Z., Deng, J. G., et al. \ 2007, \ Astronomische Nachrichten , 328, 99
\bibitem[Zhang \& Xie(2007)]{2007APSS...310...19} Zhang, Z. B., \& Xie, G. Z. \ 2007, \ Ap\&SS, 310, 19
\bibitem[Zhang et al.(2018)]{2018Nature...2...69} Zhang, B.-B., Zhang, B., Castro-Tirado, A. J., et al. \ 2018, \ Nature Astronom, 2, 69
\bibitem[Zhang et al.(2006)]{2006ChJAA...6...312} Zhang, Z.-B., Deng, J.-G., Lu, R.-J., et al. \ 2006, \ ChJA\&A, 6, 312
\end{thebibliography}
\end{document}